\documentclass[pdftex,epjc3,twocolumn]{svjour3} 
\RequirePackage{fix-cm} 
\smartqed  % flush right qed marks, e.g. at end of proof
\RequirePackage{graphicx}
\usepackage[latin1]{inputenc}     % Codepage latin1
\usepackage{mathrsfs}
\usepackage{mathtools}
\usepackage{enumitem}
\usepackage{latexsym}     % Load additional symbols
\usepackage{slashed}
\usepackage[mathscr]{eucal}     % Euler font symbols
\usepackage[normalem]{ulem}
\usepackage{graphicx}     % Include figure files, Graphics package
\usepackage{dcolumn}     % Align table columns on decimal point
\usepackage{bbm}
\usepackage{bm}     % Bold math
\usepackage[pdftex, plainpages=false, pagebackref=true, hyperindex=true, colorlinks=true, linkcolor = blue, urlcolor = blue, citecolor = blue]{hyperref}
\usepackage{placeins}
\usepackage{float}
\usepackage{subfigure}
\usepackage{helvet}
\usepackage{subfigure}
\usepackage{color,amsmath,amssymb,amsfonts,amsbsy}     % Use AMS fonts, symbols, definitions, color
\usepackage{hyperref}     % References as links
\usepackage[dvipsnames]{xcolor}
\usepackage{appendix}
\usepackage{multirow}
\usepackage{tabularx}
\usepackage{times}
\usepackage{xspace}    % Include xspace
\usepackage{rotating}
\usepackage{afterpage}
\usepackage{textcomp}
\usepackage{epsf}
\usepackage{lineno}
\usepackage{cite}
\usepackage{xurl}
\journalname{Eur. Phys. J.}

\def\empile#1\over#2{\mathrel{\mathop{\kern 0pt#1}\limits_{#2}}}

\def\beq{\begin{equation}}
\def\eeq{\end{equation}}
\def\bea{\begin{eqnarray}}
\def\eea{\end{eqnarray}}
\definecolor{darkblue}{rgb}{0.0, 0.0, 0.55}
\definecolor{darkcandyapplered}{rgb}{0.64, 0.0, 0.0}
\def\d3p{\frac{d^3\p}{(2\pi)^3}E_\p}

\newcommand{\Lb}{\left(}
\newcommand{\Rb}{\right)}

\newcommand{\gevc}{\mbox{GeV/$c$}\xspace}

\renewcommand{\thesection}{\arabic{section}}
\renewcommand{\thesubsection}{\thesection.\arabic{subsection}}
\renewcommand{\thesubsubsection}{\thesubsection.\arabic{subsubsection}}

\makeatletter
\renewcommand{\p@subsection}{}
\renewcommand{\p@subsubsection}{}
\makeatother

\newcommand{\T} {\mbox{T}}

\definecolor{darkgreen}{rgb}{0.0, 0.65, 0.0}
\definecolor{darkcyan}{rgb}{0.0, 0.65, 0.65}
\begin{document}

%%\linenumbers
%\preprint{INT-PUB-19-025}

%\title{Lowest-order QED radiative corrections beyond the ultra-relativistic limit in unpolarized 
%electron-deuteron elastic scattering for the proposed DRad experiment at Jefferson Laboratory}
\title{\centering Lowest-order QED radiative corrections in unpolarized elastic electron-deuteron scattering 
beyond the ultra-relativistic limit for the proposed deuteron charge radius measurement at Jefferson Laboratory}



\author{
Jingyi Zhou\thanksref{e1,addr1,addr2} \and \\ Vladimir Khachatryan\thanksref{e2,addr1,addr2,addr3} 
\and \\ Igor Akushevich\thanksref{e3,addr1}
\and \\ Haiyan Gao\thanksref{e4,addr1,addr2} \and \\ Alexander Ilyichev\thanksref{e5,addr4,addr5}
\and \\ Chao Peng\thanksref{e6,addr6} \and \\ Stanislav Srednyak\thanksref{e7,addr1,addr2} 
\and \\ Weizhi Xiong\thanksref{e8,addr7}    
}

\thankstext{e1}{jz271@duke.edu}
\thankstext{e2}{vladimir.khachatryan@duke.edu}
\thankstext{e3}{igor.akushevich@duke.edu}
\thankstext{e4}{haiyan.gao@duke.edu}
\thankstext{e5}{ily@hep.by}
\thankstext{e6}{cpeng@anl.gov}
\thankstext{e7}{stanislav.srednyak@duke.edu}
\thankstext{e8}{xiongw@sdu.edu.cn}

%Affiliations
\institute{Department of Physics, Duke University, Durham, NC 27708, USA \label{addr1}
\hspace{-0.8cm} \and 
Triangle Universities Nuclear Laboratory, Durham, NC 27708, USA\label{addr2}
\hspace{-0.8cm} \and 
Department of Physics, Indiana University, Bloomington, IN 47405, USA\label{addr3}
\hspace{-0.8cm} \and
Belarusian State University, Minsk, 220030, Belarus\label{addr4}
\hspace{-0.8cm} \and 
Institute for Nuclear Problems, Belarusian State University, Minsk, 220006, Belarus\label{addr5}
\hspace{-0.8cm} \and 
Physics Division, Argonne National Laboratory, Lemont, IL 60439, USA\label{addr6}
\hspace{-0.8cm} \and
Key Laboratory of Particle Physics and Particle Irradiation (MOE), Shandong University, Qingdao, Shandong 266237, China\label{addr7}
}

%\author{}
%\institute{}

\date{\today}

\maketitle

\begin{abstract}
Analogous to the well-known proton charge radius puzzle, a similar puzzle exists for the deuteron 
charge radius, $r_{d}$. There are discrepancies observed in the results of $r_{d}$, measured from
electron-deuteron ($e-d$) scattering experiments, as well as from atomic spectroscopy. 
In order to help resolve the charge radius puzzle of the deuteron, the PRad collaboration at 
Jefferson Lab has proposed an experiment for measuring $r_{d}$, named DRad. This experiment 
is designed to measure the unpolarized elastic $e-d$ scattering cross section in a low-$Q^{2}$ region.
To extract the cross section with a high precision, having reliable knowledge of QED radiative 
corrections is important. In this paper, we present complete numerical calculations of the lowest-order 
radiative corrections in $e-d$ scattering for the DRad kinematics. The calculations have been performed 
within a covariant formalism and beyond the ultra-relativistic approximation ($m_{e}^{2} \ll Q^{2}$). 
Besides, we present a systematic uncertainty on $r_{d}$ arising from higher-order radiative corrections, 
estimated based on our cross-section results.
\end{abstract}

%\tableofcontents

%%%%%%%%%%%%%%%%%%%%%%%%%%%%%%%%%%%%%%%%%%%%%%%%%%%%%%%%%%%
%%%%%%%%%%%%%%%%%%%%%%%%%%%%%%%%%%%%%%%%%%%%%%%%%%%%%%%%%%%
% For updating the reference section with these new citations
% https://inspirehep.net/literature?sort=mostrecent&size=25&page=1&q=find%20eprint%20nucl-th%2F9707016
% https://journals.aps.org/pra/pdf/10.1103/PhysRevA.56.4579

% https://arxiv.org/pdf/2206.14756.pdf
% https://arxiv.org/pdf/2206.14066.pdf
% https://arxiv.org/pdf/2203.13030.pdf
% https://arxiv.org/pdf/2212.13782.pdf
% https://arxiv.org/pdf/2109.08223.pdf
% https://arxiv.org/pdf/1711.01199.pdf

\section{\label{sec:intro} Introduction}
Studies of the internal structure of the proton and neutron, as well as their simplest bound state -- deuteron -- help 
improve our understanding of quantum chromodynamics (QCD) in the nonperturbative region. The lepton scattering experiments, 
with the availability of precisely controlled electron beams, are well-established tools for probing the nucleon
charge and magnetization distributions. In particular, if we consider the electron-proton, $e-p$
\cite{CODATA:2008,CODATA:2012,CODATA:2014,Bernauer:2010wm,Bernauer:2013tpr,Zhan:2011ji,Xiong:2019umf,Mihovilovic:2019jiz},
and electron-deuteron, $e-d$ \cite{Berard:1974ev,Simon:1981br,Platchkov:1989ch}, scattering experiments,  
the conventional proton and deuteron sizes related to their internal charge distributions are given by the 
root-mean-square (rms) charge radii, defined as 
%%%%%%%%%%%%%%%%%%%%%%%%%%%%%%%%%%%%%%%%%%%%%%%%%%%%
\bea
r_{p|d} \equiv r_{p|d,rms} & \equiv & \sqrt{\langle r_{p|d}^{2} \rangle} =
\nonumber \\
& = & \left( -6  \left. \frac{\mathrm{d} G_{E|C}^{p|d}(Q^2)}
{\mathrm{d}Q^2} \right|_{Q^{2}=0} \right)^{1/2} ,
\label{eq:eqn_rad}
\eea
%%%%%%%%%%%%%%%%%%%%%%%%%%%%%%%%%%%%%%%%%%%%%%%%%%%%
where $G_{E}^{p}$ is the proton electric form factor \cite{Ernst:1960zza,Sachs:1962zzc,Arrington:2006zm,Perdrisat:2006hj,Punjabi:2015bba}, 
$G_{C}^{d}$ -- the deuteron charge form factor 
\cite{Jankus:1957,Gourdin:1963,Garcon:2001sz,Mainz:2012,Schlimme:2016wmj,Hayward:2018qij}, and 
$Q^{2}$ -- the four-momentum transfer squared. The proton charge radius $r_{p}$ is also extracted from the laser spectroscopy 
measurements of atomic hydrogen ($e H$) and muonic hydrogen ($\mu H$). Similarly, in addition to the scattering experiments, 
the deuteron charge radius $r_{d}$ is determined from the spectroscopy measurements of atomic deuterium ($e D$) and muonic 
deuterium ($\mu D$).

In and after 2010, the two $\mu H$ spectroscopy experiments reported values of $r_{p}$ to be $0.8418 \pm 0.0007$~fm 
\cite{Pohl:2010} and $0.8409 \pm 0.0004$~fm \cite{Antognini:1900n}. The world-average value from CODATA-2014 --
$r_{p} = 0.8751 \pm 0.0061$~fm \cite{Mohr:2015ccw} -- based on the $e H$ spectroscopy measurements, along with the results 
from $e-p$ scattering experiments accomplished before 2010 mostly agree with each other, however, are larger from the 
$\mu H$ results by $\sim 7\sigma$. This discrepancy between the $r_{p}$ values, measured from those different types of 
experiments gave rise to the {\it proton charge radius puzzle} \cite{Pohl:2013yb,Carlson:2015jba,Hill:2017wzi}. 
There are a few recent $e H$ spectroscopy results, two of which \cite{Beyer:2017,Bezginov:2019} are consistent with 
the $\mu H$-based $r_{p}$ values within the measurement uncertainties.
%The recent $eH$ spectroscopy result \cite{Fleurbaey:2018} still agrees with CODATA-2014, and an $e-p$ scattering 
%experiment result \cite{Mihovilovic:2016rkr} has a large experimental uncertainty to extract definite conclusions for 
%resolving the puzzle.
The PRad collaboration at Jefferson Lab \cite{Gasparian:2014rna,Peng:2015szv} has also reported about such an agreement with 
the $\mu$H results -- $r_{p} = 0.831 \pm 0.007_{\rm stat} \pm 0.012_{\rm syst}$~fm \cite{Xiong:2019umf} -- in an unpolarized 
elastic $e-p$ scattering experiment at very low $Q^{2}$ region. With all the remarkable progress until now, there are still 
several upcoming high-precision scattering experiments to conclusively finalize the resolution of the $r_{p}$ puzzle. 
One of them is a new upgraded experiment at Jefferson Lab -- PRad-II \cite{PRad:2020oor} -- which will reduce the overall 
experimental uncertainty on $r_{p}$ by a factor of $\sim 3.8$ compared to that of PRad. 

For the $r_{p}$ determination in $e-p$ scattering experiments \cite{Gao:2021sml,Xiong:2023zih}, we know that in addition 
to knowing reliably the backgrounds associated with these experiments and having tight control of systematic uncertainties, 
it is also necessary to carefully calculate the QED radiative corrections (RCs). %Also, the interpretation of such experimental 
%data generally requires a correct calculation of RCs. 
Ref.~\cite{Akushevich:2015toa} provides analytical formulas of a complete lowest-order RC calculations for unpolarized elastic 
$e-p$ and $e-e$ (M{\o}ller) scatterings, obtained within a covariant formalism and beyond the ultra-relativistic approximation 
($m_{e}^{2} \ll Q^{2}$) for the kinematics of the PRad experiment\footnote{In the PRad experiment, most of the systematic 
uncertainties are correlated and $Q^{2}$-dependent}. The contribution to the total systematic uncertainty from RCs arise 
mostly from their higher-order contributions to the cross section \cite{Arbuzov:2015vba}. If the cross-section uncertainties 
are turned into uncertainties on $r_{p}$, then the absolute RC systematic uncertainty in PRad for $e-p$ due to higher-order 
RCs is $0.0020$~fm, whereas it is $0.0065$~fm for the M{\o}ller process \cite{Xiong:2019umf}. The total RC systematic uncertainty 
is $\Delta r_{p} = 0.0069$~fm.

Interestingly enough, there is also the {\it deuteron charge radius puzzle} in determination of $r_{d}$ in $e-d$ scattering 
experiments and $e D$\big/$\mu D$ spectroscopy measurements. Since 1998, the $r_{d}$ input (determined from elastic $e-d$ 
scattering data) in the CODATA adjustment is $r_{d} = 2.13 \pm 0.01$~fm, obtained in \cite{Trautmann:1998,Sick:2001}. 
Meanwhile, the most recent scattering-based $r_{d}$ input in the CODATA-2018 adjustment \cite{Tiesinga:2021myr} is 
$r_{d} = 2.111 \pm 0.019$~fm. In 2016, the CREMA collaboration has reported a deuteron charge radius value -- 
$r_{d} = 2.12562 \pm 0.00078$~fm -- from a $\mu D$-based spectroscopy measurement of three $2P \rightarrow 2S$ transitions 
in $\mu D$ atoms \cite{Pohl:2016}, which is 3.2$\sigma$ smaller than the CODATA-2018 world-average value:
$r_{d} = 2.12799 \pm 0.00074$~fm \cite{Tiesinga:2021myr}. On the other hand, the radius from \cite{Pohl:2016} is 
3.5$\sigma$ smaller than $r_{d} = 2.1415 \pm 0.0045$~fm determined from an $e D$-based measurement \cite{Pohl:2016glp} 
of $1S \rightarrow 2S$ transitions in $e D$ atoms, along with these transitions measured in \cite{Parthey:2010aya}.

The uncertainties from all previous $e-d$ scattering experiments are too large to contribute to a satisfactory resolution 
of the $r_{d}$ puzzle. In this respect, the PRad collaboration has proposed a new high-precision magnetic-spectrometer-free,
calorimeter-based unpolarized elastic $e-d$ scattering experiment (named DRad \cite{DRad}) at the scattering angle range 
$\theta_{e} = 0.7^{\circ} - 6.0^{\circ}$, and at electron beam energies $E_{1} = 1.1~{\rm GeV}$ 
and $2.2~{\rm GeV}$, which corresponds to $Q^{2} = 2\times 10^{-4}~(\gevc)^{2}$ $-$ $5\times 10^{-2}~(\gevc)^{2}$. 
The DRad experiment is designed to utilize the PRad-II experimental setup, by adding a low-energy 
cylindrical recoil detector for ensuring the elasticity of the $e-d$ scattering process. The proposed experiment will allow 
a model-independent extraction of $r_{d}$ with a precision $\sim0.2\%$ for addressing the $r_{d}$ puzzle. One should note 
that the M{\o}ller scattering will be simultaneously measured in the DRad experiment, making use of the same procedure and 
tools that have been deployed in PRad, 
%and which will also be applied to PRad-II, 
for controlling the systematic uncertainties
associated with the absolute $e-p$ and $e-d$ cross-section measurements along with monitoring the luminosity. Moreover, it is
similarly necessary to carefully calculate the QED RCs for this experiment. For this purpose, we will use the RC covariant 
framework developed in 
\cite{Akushevich:2015toa,Akushevich:1994dn,Akushevich:2001yp,Afanasev:2001jd,Akushevich:2019mbz,Ilyichev:2005rx,Akushevich:2007jc,Zykunov:2021fxh,Afanasev:2021nhy,Afanasev:2022uwm} 
to perform calculations on the lowest-order Feynman diagrams for unpolarized elastic $e-d$ scattering\footnote{For 
alternative lowest- and also higher-order RC calculations and event generators, we refer to
Refs.~\cite{Maximon:2000hm,Gramolin:2014pva,Bucoveanu:2018soy,Bucoveanu:2019hxz,Fadin:2018dwp,Banerjee:2020rww,Kaiser:2022pso,Schmookler:2022gxw}
in unpolarized lepton-proton scattering and to Refs.~\cite{Banerjee:2020rww,Shumeiko:1999zd,Kaiser:2010zz,Epstein:2016lpm,Aleksejevs:2013,Aleksejevs:2020vxy,Banerjee:2021mty,Banerjee:2021qvi,Bondarenko:2022kxq} 
in lepton-lepton scattering.}. 
In our final results the electron mass $m_{e}$ is taken into account. In this framework, the Bardin-Shumeiko approach 
is used for the covariant extraction and cancellation of infrared divergences \cite{Bardin:1976qa,Shumeiko:1978cn}. 
At this point, we wish to emphasize that the differential cross section of elastic scattering of deuterons on electrons at rest
has been studied theoretically with great details taking model-independent QED RCs into account \cite{Gakh:2018sat}, in which 
a high-energy incident deuteron and a recoil electron are both considered to be detected in a coincidence experimental setup, 
resulting in kinematic coverages at very low $Q^{2}$. 

It is also relevant to indicate some recent phenomenological and theoretical developments related to the deuteron radius 
extraction. Like Ref.~\cite{Zhou:2020cdt}, in which a robust extraction of $r_{d}$ prior to the DRad data taking is carried 
out, based on the ``root mean square error (RMSE)" method developed by the PRad collaboration in \cite{Yan:2018bez}.
Ref.~\cite{Cui:2022fyr} shows an alternative robust extraction of $r_{d}$ %again prior to anticipated DRad data taking, {\color{cyan} they didn't meant to do it for DRad}
using the ``statistical Schlessinger point method (SPM)" first applied to the $r_{p}$ extraction by Craig Roberts and his 
collaborators in \cite{Cui:2021vgm}. Refs.~\cite{Filin:2019eoe,Filin:2020tcs} present high-accuracy calculations of the 
deuteron radius in chiral effective field theory, along with the determination of the neutron charge radius, by Evgeny Epelbaum
and his collaborators.

Our paper is organized as follows. In Sec.~\ref{sec:ff_models}, we discuss the unpolarized elastic $e-d$ differential 
cross section. A short overview of the deuteron four data-based electromagnetic form-factor models (with pertinent details) 
is given in Appendix A because the final cross sections include the form-factor parametrizations of the deuteron. 
In Sec.~\ref{sec:xs}, we first discuss the kinematics of the process, and then discuss in detail the model-independent 
lowest-order RCs in $e-d$, including corrections stemming from a lepton vertex function, vacuum polarization, and 
radiation of a real photon from leptonic legs. Sec.~\ref{sec:numeric} presents numerical results on the computed final 
cross sections, and estimation of the lowest-order RC systematic uncertainty on $r_{d}$ for the DRad experiment. 
We summarize and provide prospects in the last section.

%%%%%%%%%%%%%%%%%%%%%%%%%%%%%%%%%%%%%%%%%%%%%%%%%%%%%%%%%%%
%%%%%%%%%%%%%%%%%%%%%%%%%%%%%%%%%%%%%%%%%%%%%%%%%%%%%%%%%%%
\section{\label{sec:ff_models} Electron-deuteron unpolarized elastic cross section and 
deuteron electromagnetic form-factor models}

One of the fundamental problems in modern nuclear physics is to understand the deuteron's electromagnetic structure,
given that it is the only naturally-existing two-nucleon bound system. Elastic scattering in the $e-d$ process is 
more complicated than that in the $e-p$ process, however, it is anticipated that at low $Q^{2}$ the deuteron form 
factors are controlled by part of its wave function, for which the two nucleons are quite apart. This is because 
the non-nucleonic degrees of freedom and relativistic effects inside the deuteron are expected to be insignificant. 
As a consequence, theoretical computations of the deuteron form factors and rms radius are considered to be well 
grounded since those are independent of a broad class of nucleon-nucleon potentials, and depend mostly on the 
neutron-proton scattering length and their binding energy \cite{Wong:1994sy}. This makes, e.g., $r_{d}$ an ideal 
observable for theory-experiment comparisons. 

\begin{figure}[hbt]
%\vspace{-5mm}
\centering
\includegraphics[width=4.15cm]{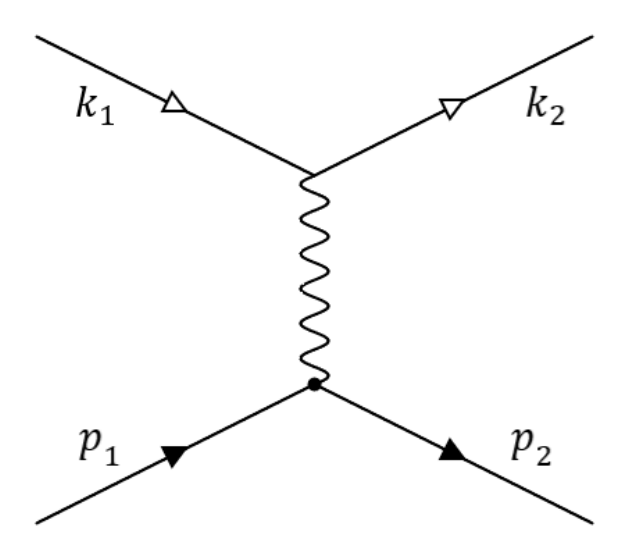}
\caption{Feynman diagram contributing to the Born cross section for the elastic $e-d$ scattering. Only the four-momenta
of the electron and deuteron are shown.}
\label{fig:fig_feynman_born}
\end{figure}

In the general case of the $e-d$ scattering process, the four-momentum and helicity of incident\,$\&$\,scattered electrons
off a spin-1 deuteron fixed target are $(k_{1},\sigma_{1})$\,$\&$\,$(k_{2},\sigma_{2})$, respectively, and the corresponding 
quantities for the target are $(p_{1},\lambda_{1})$\,$\&$\,$(p_{2},\lambda_{2})$ (see Fig.~\ref{fig:fig_feynman_born}).
%-----------------------------------------Fig1-----------------------------------------
%---------------------------------------------------------------------------------------

The matrix element of the electromagnetic current operator for this process has the following form 
\cite{Arnold:1980zj,Garcon:2001sz}:
%%%%%%%%%%%%%%%%%%%%%%%%%%%%%%%%%%%%%%%%%%%%%%%%%%%%
\bea
& & \mathcal{M} = 
\nonumber \\
& & ~~~~~= i e^{2} \bar{u}(k_{2},\sigma_{2})\gamma^{\mu}u(k_{1},\sigma_{1})
\frac{1}{q^{2}} \langle p_{2},\lambda_{2}|j_{\mu}|p_{1},\lambda_{1} \rangle ,
\label{eq:eqn_matrix}
\eea
%%%%%%%%%%%%%%%%%%%%%%%%%%%%%%%%%%%%%%%%%%%%%%%%%%%%
where $\bar{u}$ and $u$ are the Dirac spinors, $q^{2} = \Lb p_{2} - p_{1} \Rb^{2} = \Lb k_{1} - k_{2} \Rb^{2}$ is the 
four-momentum transfer squared carried by the exchanged virtual photon. The electromagnetic current operator for the 
deuteron is given by
%%%%%%%%%%%%%%%%%%%%%%%%%%%%%%%%%%%%%%%%%%%%%%%%%%%%
\bea
& & \!\!\!
\langle p_{2},\lambda_{2}|j_{\mu}|p_{1},\lambda_{1} \rangle \equiv G_{\lambda_{2},\lambda_{1}}^{\mu}(p_{2},p_{1}) =
\nonumber \\
& & = - \bigg\{ G_{1}^{d}(q^{2}) \Lb \xi_{\lambda_{2}}^{\ast}(p_{2}) \cdot \xi_{\lambda_{1}}(p_{1}) \Rb \Lb p_{2} + p_{1} \Rb^{\mu} 
\bigg. +
\nonumber \\
& & ~~~~~~~~~~
+ G_{2}^{d}(q^{2}) \bigg[ \xi_{\lambda_{1}}^{\mu}(p_{1}) \Lb \xi_{\lambda_{2}}^{\ast}(p_{2}) \cdot q \Rb - \bigg.
\nonumber \\
& & ~~~~~~~~~~~~~~~~~~~~~~~~~~~~~~~~~~~~~~~~
\bigg. - \xi_{\lambda_{2}}^{\mu\ast}(p_{2}) \Lb \xi_{\lambda_{1}}(p_{1}) \cdot q \Rb \bigg] -
\nonumber \\
& & ~~~~~~~~~~
- G_{3}^{d}(q^{2}) \frac{1}{2M_{d}^{2}} \Lb \xi_{\lambda_{2}}^{\ast}(p_{2}) \cdot q \Rb \Lb \xi_{\lambda_{1}}(p_{1}) \cdot q \Rb \times
\nonumber \\
& & ~~~~~~~~~~~~~~~~~~~~~~~~~~~~~~~~~~~~~~~~~~~~~~~~~~~~~~~~~~~~
\bigg. \times \Lb p_{2} + p_{1} \Rb^{\mu} \bigg\} ,
\label{eq:eqn_current}
\eea
%%%%%%%%%%%%%%%%%%%%%%%%%%%%%%%%%%%%%%%%%%%%%%%%%%%%
where $M_{d}$ is the deuteron mass (= 1875.612~MeV). $\xi_{1} \equiv \xi_{\lambda_{1}}^{\mu}(p_{1})$ and 
$\xi_{2} \equiv \xi_{\lambda_{2}}^{\mu\ast}(p_{2})$ are the polarization four-vectors of the initial and final 
deuteron states, satisfying the condition of $\xi_{1} \cdot p_{1} = \xi_{2} \cdot p_{2}$. 

For electron scattering, since the virtual photon four-momentum is always space-like, the convention 
$Q^{2} \equiv -q^{2}$ is adopted. The $Q^{2}$-dependent form factors $G_{i}^{d}$ in Eq.~(\ref{eq:eqn_current}) 
are related to the charge monopole ($G_{C}^{d}$), magnetic dipole ($G_{M}^{d}$), and charge quadrupole 
($G_{Q}^{d}$) form factors via
%%%%%%%%%%%%%%%%%%%%%%%%%%%%%%%%%%%%%%%%%%%%%%%%%%%%
\bea
G_{C}^{d}(Q^{2}) & = & G_{1}^{d}(Q^{2}) + \frac{2}{3}\,\eta\,G_{Q}^{d}(Q^{2}) ,
\nonumber \\
G_{M}^{d}(Q^{2}) & = & G_{2}^{d}(Q^{2}) ,
\nonumber \\
G_{Q}^{d}(Q^{2}) & = & G_{1}^{d}(Q^{2}) - G_{2}^{d}(Q^{2}) + (1 + \eta)\,G_{3}^{d}(Q^{2}) ,
\label{eq:eqn_Gi}
\eea
%%%%%%%%%%%%%%%%%%%%%%%%%%%%%%%%%%%%%%%%%%%%%%%%%%%%
with $\eta = Q^{2}/\Lb 4M_{d}^{2} \Rb$. Besides, there are the following additional relations that are normalized such that
%%%%%%%%%%%%%%%%%%%%%%%%%%%%%%%%%%%%%%%%%%%%%%%%%%%%
\beq
G_{C}^{d}(0) = 1 ,\,\,\,\,\,\,\,\,\frac{G_{M}^{d}(0)}{\mu_{M}^{d}} = 1 ,\,\,\,\,\,\,\,\frac{G_{Q}^{d}(0)}{\mu_{Q}^{d}} = 1 ,
\label{eq:eqn_G0}
\eeq
%%%%%%%%%%%%%%%%%%%%%%%%%%%%%%%%%%%%%%%%%%%%%%%%%%%%
with the given deuteron magnetic dipole moment ($\mu_{M}^{d}$) and electric quadrupole moment 
($\mu_{Q}^{d}$)\footnote{Throughout our work, we use dimensionless quantities $G_{M}^{d}(0) = 1.714$, 
$\mu_{M}^{d} \equiv (\mu_{M}^{d}/\mu_{N}$) = 0.8574, and 
$G_{Q}^{d}(0) = 25.830$, $\mu_{Q}^{d} \equiv (\mu_{Q}^{d}/{\rm fm^{2}})$ = 0.2859 \cite{Garcon:2001sz}.}.

The cross section for elastic scattering of longitudinally polarized electrons off a polarized deuteron target 
is calculated in the rest frame of the deuteron, within the Born approximation assuming one-photon exchange 
\cite{Arnold:1980zj,Donnelly:1985ry,Garcon:2001sz}. Nonetheless, the electron beam and the deuteron target in the 
DRad experiment are planned to be unpolarized. Therefore, we are interested in the process of
%%%%%%%%%%%%%%%%%%%%%%%%%%%%%%%%%%%%%%%%%%%%%%%%%%%
\beq
e(k_{1}) + d(p_{1}) \rightarrow e^{\prime}(k_{2}) + d(p_{2}) ,
\label{eqn_eq:reaction}
\eeq
%%%%%%%%%%%%%%%%%%%%%%%%%%%%%%%%%%%%%%%%%%%%%%%%%%%
and in using the following well-known Born cross section of the unpolarized elastic $e-d$ scattering at low $Q^{2}$
\cite{Berard:1974ev,Simon:1981br,Platchkov:1989ch,Coester:1975hj,Arnold:1980zj,Abbott:2000ak,kobushkin1995deuteron}:
%%%%%%%%%%%%%%%%%%%%%%%%%%%%%%%%%%%%%%%%%%%%%%%%%%%
\bea
& & \frac{\mathrm{d}\sigma^{B}}{\mathrm{d}\theta_{e}}\!\Lb E_{1}, \theta_{e} \Rb  = \sigma_{_{\!NS}}\!\Lb E_{1}, \theta_{e} \Rb \times
\nonumber \\
& & ~~~~~~~~~~~~~~~~~~~~~~
\times \Lb A_{d}(Q^{2}) + B_{d}(Q^{2})\,\tan^{2}{\!\Lb \frac{\theta_{e}}{2} \Rb} \Rb .
\label{eq:eqn_Born}
\eea
%%%%%%%%%%%%%%%%%%%%%%%%%%%%%%%%%%%%%%%%%%%%%%%%%%%%
Here the unpolarized elastic structure functions $A_{d}(Q^{2})$ and $B_{d}(Q^{2})$ are defined as
%%%%%%%%%%%%%%%%%%%%%%%%%%%%%%%%%%%%%%%%%%%%%%%%%%%%
\bea
A_{d}(Q^{2}) & = & \Lb G_{C}^{d}(Q^{2}) \Rb^{2} + \frac{2}{3}\,\eta \Lb G_{M}^{d}(Q^{2}) \Rb^{2} +
\nonumber \\
& & + \frac{8}{9}\,\eta^{2} \Lb G_{Q}^{d}(Q^{2}) \Rb^{2} ,
\nonumber \\
B_{d}(Q^{2}) & = & \frac{4}{3}\,\eta (1 + \eta) \Lb G_{M}^{d}(Q^{2}) \Rb^{2} ,
\label{eq:eqn_deuteronstrucfunc}
\eea
%%%%%%%%%%%%%%%%%%%%%%%%%%%%%%%%%%%%%%%%%%%%%%%%%%%%
and $\sigma_{_{\!NS}}$ is the differential Mott cross section for the elastic scattering from a point-like and spinless 
particle at the electron scattering angle $\theta_{e}$ and the incident energy $E_{1}$ \cite{Bernauer:2010zga}: 
%%%%%%%%%%%%%%%%%%%%%%%%%%%%%%%%%%%%%%%%%%%%%%%%%%%%
\bea
& & \sigma_{_{\!NS}}\!\Lb E_{1}, \theta_{e} \Rb \equiv \Lb \frac{\mathrm{d}\sigma}{\mathrm{d}\theta_{e}}\!\Lb E_{1}, 
\theta_{e} \Rb \Rb_{\rm Mott} =
\nonumber \\
& & ~~~~~~~~
= 2\pi\,\frac{4\alpha^{2} E_{2}^{2}}{Q^{4}}\frac{E_{2}}{E_{1}} \Lb 1 - \sin^{2}{\!\Lb \frac{\theta_{e}}{2} \Rb} \Rb \sin{\!(\theta_{e})} .
\label{eq:eqn_Mott}
\eea
%%%%%%%%%%%%%%%%%%%%%%%%%%%%%%%%%%%%%%%%%%%%%%%%%%%%
where $\alpha = e^{2}/4\pi = 1/137.036$ is the electromagnetic fine-structure constant.
It should be mentioned that in the literature the cross sections in Eq.~(\ref{eq:eqn_Born}) and Eq.~(\ref{eq:eqn_Mott}) 
are usually given as $\mathrm{d}\sigma/\mathrm{d}\Omega$. Meanwhile, we use the relation 
$\mathrm{d}\Omega = \sin{\!(\theta_{e})}\,\mathrm{d}\theta_{e}\,\mathrm{d}\varphi$,
and uniformly integrate the cross section over the azimuthal angle $\varphi$ from 0 to $2\pi$ limits\footnote{In Appendix B, 
we show $Q^{2}$-dependent Born cross section (equivalent to Eq.~(\ref{eq:eqn_Born})) derived in the ansatz of 
\cite{Akushevich:2015toa,Afanasev:2021nhy}.}.

Generally, if we consider the lepton-deuteron scattering, its energy conservation reads as \cite{Greiner:2009}
%%%%%%%%%%%%%%%%%%%%%%%%%%%%%%%%%%%%%%%%%%%%%%%%%%%%
\beq
M_{d} \Lb E_{1} - E_{2} \Rb = E_{1}E_{2} - |\textit{\textbf{p}$_{1}$}||\textit{\textbf{p}$_{2}$}| \cos{\!(\theta_{l})} - m_{l}^{2} .
\label{eq:eqn_cons}
\eeq
%%%%%%%%%%%%%%%%%%%%%%%%%%%%%%%%%%%%%%%%%%%%%%%%%%%%
The scattered lepton energy $E_{2}$ is solved to be
%%%%%%%%%%%%%%%%%%%%%%%%%%%%%%%%%%%%%%%%%%%%%%%%%%%%
\beq
E_{2} = \frac{\mathcal{B} + \sqrt{\mathcal{B}^{2} + 4\mathcal{A}\,\mathcal{C}}}{-2\mathcal{A}} ,
\label{eq:eqn_E2}
\eeq
%%%%%%%%%%%%%%%%%%%%%%%%%%%%%%%%%%%%%%%%%%%%%%%%%%%%
where
%%%%%%%%%%%%%%%%%%%%%%%%%%%%%%%%%%%%%%%%%%%%%%%%%%%%
\bea
& & 
\mathcal{A} = \Lb E_{1}^{2} - m_{l}^{2} \Rb \cos^{2}{\!(\theta_{l})} - E_{1}^{2} - 2M_{d}E_{1} - M_{d}^{2} ,
\nonumber \\
& &
\mathcal{B} = 2 \Lb m_{l}^{2} + M_{d}^{2} \Rb E_{1} + 2M_{d} \Lb m_{l}^{2} + E_{1}^{2} \Rb ,
\nonumber \\
& &
\mathcal{C} = \Lb m_{l}^{2}E_{1}^{2} - m_{l}^{4} \Rb \cos^{2}{\!(\theta_{l})} + 
\nonumber \\
& & ~~~~~~~
+ m_{l}^{4} + 2m_{l}^{2}M_{d}E_{1} + M_{d}^{2}E_{1}^{2} .
\label{eq:eqn_coeff}
\eea
%%%%%%%%%%%%%%%%%%%%%%%%%%%%%%%%%%%%%%%%%%%%%%%%%%%%
In the limit of $E_{1}/m_{l}$, $E_{2}/m_{l}$, $M_{d}/m_{l} \gg 1$ (which is the case for the $e-d$ scattering in the DRad 
kinematics), Eq.~(\ref{eq:eqn_E2}) reduces to
%%%%%%%%%%%%%%%%%%%%%%%%%%%%%%%%%%%%%%%%%%%%%%%%%%%%
\beq
E_{2} = \frac{E_{1}}{1 + (2E_{1}/M_{d}) \sin^{2}(\!\theta_{l}/2)} .
\label{eq:eqn_E2b}
\eeq
%%%%%%%%%%%%%%%%%%%%%%%%%%%%%%%%%%%%%%%%%%%%%%%%%%%%
Another relation between the momentum transfer squared and the scattering angle is
%%%%%%%%%%%%%%%%%%%%%%%%%%%%%%%%%%%%%%%%%%%%%%%%%%%%
\beq
q^{2} = 2 \Lb m_{l}^{2} - E_{1}E_{2} + |\textit{\textbf{p}$_{1}$}||\textit{\textbf{p}$_{2}$}| \cos{\!(\theta_{l})} \Rb ,
\label{eq:eqn_q2_theta}
\eeq
%%%%%%%%%%%%%%%%%%%%%%%%%%%%%%%%%%%%%%%%%%%%%%%%%%%%
which brings to
%%%%%%%%%%%%%%%%%%%%%%%%%%%%%%%%%%%%%%%%%%%%%%%%%%%%
\beq
\cos{\!(\theta_{l})} = \frac{2E_{1}E_{2} - 2m_{l}^{2} - Q^{2}}{2\sqrt{E_{1}^{2} - m_{l}^{2}}\,\sqrt{E_{2}^{2} - m_{l}^{2}}} ,
\label{eq:eqn_theta}
\eeq
%%%%%%%%%%%%%%%%%%%%%%%%%%%%%%%%%%%%%%%%%%%%%%%%%%%%
If one neglects the lepton mass, it reduces to another well-known formula of $Q^{2} = 4E_{1}E_{2}\sin^{2}(\!\theta_{l}/2)$, 
however, as it is mentioned in the introduction, the DRad experiment will reach low $Q^{2} \sim 10^{-4}~(\gevc)^{2}$ region,
where it is necessary to keep $m_{l}$ in Eq.~(\ref{eq:eqn_theta}). The lowest-order RC results and the total (observed) 
cross section are in particular obtained in Sec.~\ref{sec:xs}, with $m_{l}$ taken in calculations into account that is 
beyond the ultra-relativistic approximation.
 
On the other hand, Eq.~(\ref{eq:eqn_Born}) has been derived in literature by neglecting the electron (lepton) mass, however, 
its inclusion into that formula would not affect the cross section determination with very high accuracy, at JLab and Mainz 
beam energies. The reason is that in the extraction of $r_{p}$ \cite{Xiong:2019umf,Xiong:2020kds}, based on using a similar Born 
cross-section formula for unpolarized elastic $e-p$ scattering at low $Q^{2}$, there were essentially no changes in the 
$r_{p}$ central value and its total systematic uncertainty, with and without the electron mass included in the Born 
cross section. Furthermore, the massless Born formula in Eq.~(\ref{eq:eqn_Born}) has been used by the Mainz Microtron A1 
collaboration for the deuteron form-factor precise measurements in the elastic $e-d$ scattering \cite{Mainz:2012,Schlimme:2016wmj}.

%%%%%%%%%%%%%%%%%%%%%%%%%%%%%%%%%%%%%%%%%%%%%%%%%%%%%%%%%%%
%%%%%%%%%%%%%%%%%%%%%%%%%%%%%%%%%%%%%%%%%%%%%%%%%%%%%%%%%%%
\section{\label{sec:xs} Lowest-order QED radiative corrections to the unpolarized elastic $e-d$ scattering cross section}

In this section, we deploy some of the key formulas and derivations from 
Refs.~\cite{Afanasev:2021nhy,Akushevich:2015toa} (partially used in the ansatz of 
\cite{Akushevich:1994dn,Akushevich:2019mbz,Byer:2022bqf} too) since those expressions can also be used for the unpolarized 
elastic $e-d$ scattering, as far as the DRad experiment is concerned. Meanwhile, one should mention that all the RC and 
cross-section formulas represented here are obtained within the covariant Bardin-Shumeiko formalism 
\cite{Bardin:1976qa,Shumeiko:1978cn}, which is applied to the extraction and cancellation of infrared divergences stemming 
from virtual and real photon radiation in the scattering process under consideration. The ultimate results derived 
within this formalism are independent of any unphysical and artificial parameters, like the cut-off parameter existing 
in the Mo-Tsai approach \cite{Mo:1968cg,Tsai:1971qi}, introduced for separation of the regions of soft and hard photon 
radiation, while canceling out the infrared divergences in the pertinent Feynman diagram calculations.

%%%%%%%%%%%%%%%%%%%%%%%%%%%%%%%%%%%%%%%%%%%%%%%%%%%%%%%%%%%
\subsection{\label{sec:tensor} Kinematics of the Bremsstrahlung process}

\begin{figure}[h!]
\centering
\includegraphics[width=4.15cm]{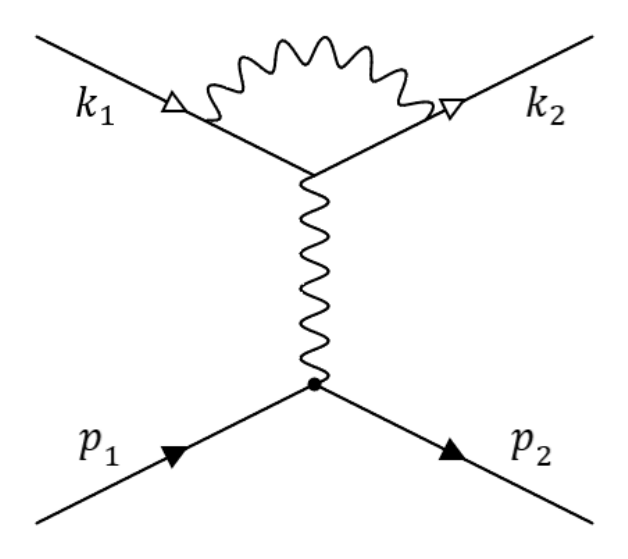}\label{fig:fig_feynman_rc1}
\includegraphics[width=4.15cm]{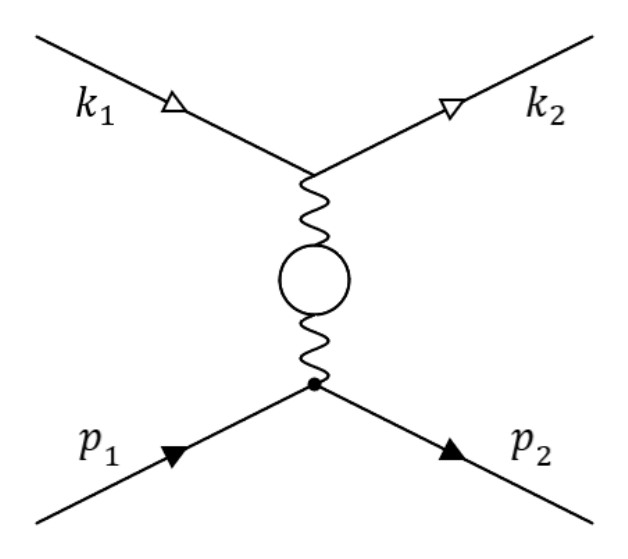}\label{fig:fig_feynman_rc2}
\\[-0.1cm]
{\bf (a) \hspace{3.75cm} (b)}
\\[0.1cm]
\vspace{3mm}
\includegraphics[width=4.15cm]{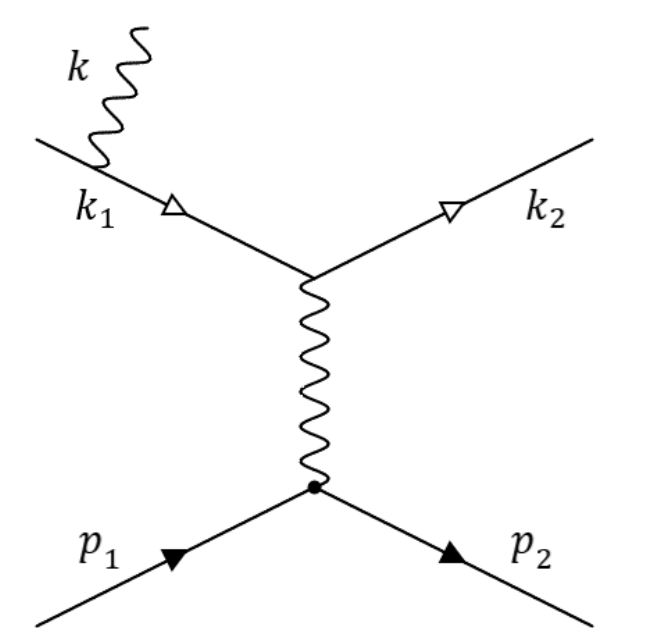}\label{fig:fig_feynman_rc3}
\includegraphics[width=4.15cm]{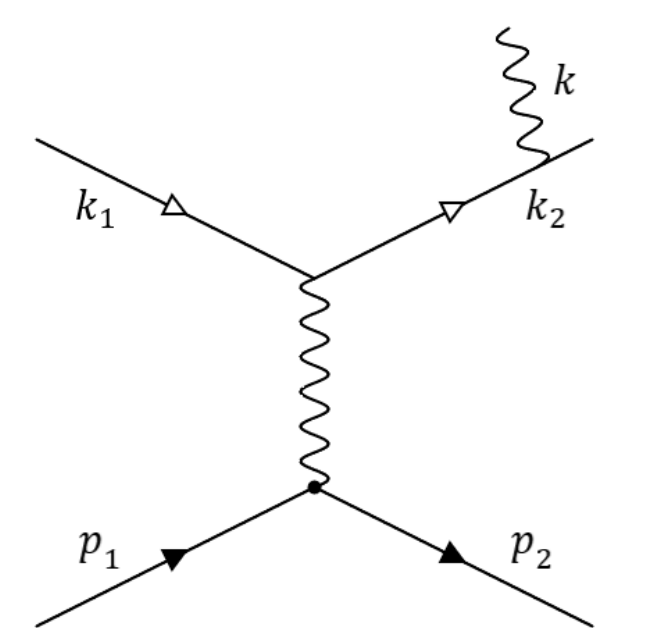}\label{fig:fig_feynman_rc4}
\\[-0.1cm]
{\bf (c) \hspace{3.75cm} (d)}
\\[-0.1cm]
\vspace{2mm}
\caption{Feynman diagrams from (a) to (d), describing the lowest-order QED RC contributions to the unpolarized elastic 
$e-d$ scattering cross section: (a) vertex correction; (b) vacuum polarization; (c), (d) electron-leg bremsstrahlung.} 
\label{fig:fig_feynman_rcs}
\end{figure}

The Feynman diagram of the Born process is already shown in Fig.~\ref{fig:fig_feynman_born}. The other diagrams of interest 
for computations of the lowest-order RCs are demonstrated in Fig.~\ref{fig:fig_feynman_rcs}, which show the vertex correction (a), 
the vacuum polarization (b), as well as the bremsstrahlung (c) and (d) originating from the leptonic legs.
%-----------------------------------------Fig2-----------------------------------------
%---------------------------------------------------------------------------------------
Let us start by considering the process of bremsstrahlung with a radiated hard photon, $\gamma(k)$, represented as
%%%%%%%%%%%%%%%%%%%%%%%%%%%%%%%%%%%%%%%%%%%%%%%%%%%
\beq
e(k_{1}) + d(p_{1}) \rightarrow e^{\prime}(k_{2}) + d(p_{2}) + \gamma(k) ,
\label{eqn_eq:brem}
\eeq
%%%%%%%%%%%%%%%%%%%%%%%%%%%%%%%%%%%%%%%%%%%%%%%%%%%
the cross section of which is given by
%%%%%%%%%%%%%%%%%%%%%%%%%%%%%%%%%%%%%%%%%%%%%%%%%%%
\beq
\mathrm{d}\sigma_{R} = \frac{1}{2\sqrt{\lambda_{S}}}\,\mathcal{M}_{R}^{2}\,\mathrm{d}\Gamma_{3} , 
\label{eqn_eq:brem_xs}
\eeq
%%%%%%%%%%%%%%%%%%%%%%%%%%%%%%%%%%%%%%%%%%%%%%%%%%%
where
%%%%%%%%%%%%%%%%%%%%%%%%%%%%%%%%%%%%%%%%%%%%%%%%%%%
\beq
\lambda_{S} = S^{2} - 4m_{e}^{2}M_{d}^{2},~~~~\mbox{with}~~~S = 2k_{1}\!\cdot\!p_{1} = 2E_{1}\!\cdot\!M_{d} .
\label{eqn_eq:lambdaS}
\eeq
%%%%%%%%%%%%%%%%%%%%%%%%%%%%%%%%%%%%%%%%%%%%%%%%%%%
For the matrix element squared $\mathcal{M}_{R}^{2}$, we refer to the equation (49) in \cite{Afanasev:2021nhy}, and the equation
(14) in \cite{Akushevich:2015toa}. The phase-space element $\mathrm{d}\Gamma_{3}$ can be described in terms of three additional 
quantities (``photonic variables"). For such an expression of the phase space, we refer to equation (48) in 
\cite{Afanasev:2021nhy}. We choose the standard set of photonic variables to be
%%%%%%%%%%%%%%%%%%%%%%%%%%%%%%%%%%%%%%%%%%%%%%%%%%%
\beq
\upsilon = (p_{1} + k_{1} - k_{2})^{2} - M_{d}^{2},~~~~~\tau = \frac{k\!\cdot\!q}{k\!\cdot\!p_{1}} ,~~~~~\phi_{k} ,
\label{eqn_eq:3var}
\eeq
%%%%%%%%%%%%%%%%%%%%%%%%%%%%%%%%%%%%%%%%%%%%%%%%%%%
in which $\upsilon$ is the inelasticity, and $\phi_k$ is the azimuthal angle between (${\bf k}_1$, ${\bf k}_2$) and (${\bf k}$,
${\bf q}$) planes in the rest frame (${\bf p}_{1}=0$). The upper limit for $\upsilon$ at fixed $Q^{2}$ is
%%%%%%%%%%%%%%%%%%%%%%%%%%%%%%%%%%%%%%%%%%%%%%%%%%%
\beq
\upsilon_{q} = \frac{\sqrt{\lambda_{S}}\sqrt{\lambda_{m}} - Q^{2}\Lb S + 2m_{e}^{2}\Rb}{2m_{e}^{2}} .
\label{eqn_eq:vq}
\eeq
%%%%%%%%%%%%%%%%%%%%%%%%%%%%%%%%%%%%%%%%%%%%%%%%%%%
where the function $\lambda_{m}$ is given by
%%%%%%%%%%%%%%%%%%%%%%%%%%%%%%%%%%%%%%%%%%%%%%%%%%%%%%%%%%%%%%%%%%%
\beq
\lambda_{m} = Q^{2} \Lb Q^{2} + 4m_{e}^{2} \Rb ,
\label{eqn_eq:lambda_m}
\eeq
%%%%%%%%%%%%%%%%%%%%%%%%%%%%%%%%%%%%%%%%%%%%%%%%%%%%%%%%%%%%%%%%%%%
When $Q^{2}$ is near its kinematic boundaries, the maximum value of the inelasticity $\upsilon_{q}$ is determined to be
%%%%%%%%%%%%%%%%%%%%%%%%%%%%%%%%%%%%%%%%%%%%%%%%%%%
\beq
\upsilon_{q}^{\rm max} = S - 2m_{e}\!\left( \sqrt{S + m_{e}^{2} + M_{d}^{2}} - m_{e} \right) .
\label{eqn_eq:vqmax}
\eeq
%%%%%%%%%%%%%%%%%%%%%%%%%%%%%%%%%%%%%%%%%%%%%%%%%%%
In practice, the real hard photon contribution to the observed cross section can be considerably reduced by applying a cut 
on the inelasticity quantity. This cut in turn is a measured quantity in single-arm measurements of elastically scattered leptons only. 
Keeping in mind the maximum value $\upsilon_{q}^{\rm max}$, throughout this section we use an experimentally observable variable, 
$\upsilon_{\rm cut}$, for the upper limit of the inelasticity (both for the fixed $Q^{2}$ and $\theta_{e}$). 
Sec.~\ref{sec:numeric_rc} discusses more details on $\upsilon_{\rm cut}$.

%%%%%%%%%%%%%%%%%%%%%%%%%%%%%%%%%%%%%%%%%%%%%%%%%%%%%%%%%%%
\subsection{\label{sec:rc} Lowest-order radiative corrections and the observed cross section}

We use Eqs.~(\ref{eq:eqn_Born}), (\ref{eq:eqn_deuteronstrucfunc}), (\ref{eq:eqn_Mott}), and (\ref{eq:eqn_E2b}) for the unpolarized 
elastic $e-d$ Born cross section in our current analysis. The equivalent cross section is also shown in Appendix B. 
The $Q^{2} \leftrightarrow \theta_{e}$ transformation is given by Eq.~(\ref{eq:eqn_theta}), otherwise expressed as 
%%%%%%%%%%%%%%%%%%%%%%%%%%%%%%%%%%%%%%%%%%%%%%%%%%%%
\bea
Q^{2} & = & 2E_{1}E_{2} - 2m_{e}^{2} - 
\nonumber \\
& & - 2\sqrt{E_{1}^{2} - m_{e}^{2}}\,\sqrt{E_{2}^{2} - m_{e}^{2}}\,\cos{\!(\theta_{e})} ,
\label{eq:eqn_Q2_costheta}
\eea
%%%%%%%%%%%%%%%%%%%%%%%%%%%%%%%%%%%%%%%%%%%%%%%%%%%%
On the other hand, from Ref.~\cite{Afanasev:2021nhy} we have
%%%%%%%%%%%%%%%%%%%%%%%%%%%%%%%%%%%%%%%%%%%%%%%%%%%%%%%%%%%%%%%%%%%
\beq
\frac{\mathrm{d}\sigma^{B}}{\mathrm{d}Q^{2}} = -\frac{1}{j_{\theta}\sin{\!(\theta_{e})}}
\frac{\mathrm{d}\sigma^{B}}{\mathrm{d}\theta_{e}} ,
\label{eqn_eq:dxs_Born_dQ2}
\eeq
%%%%%%%%%%%%%%%%%%%%%%%%%%%%%%%%%%%%%%%%%%%%%%%%%%%%%%%%%%%%%%%%%%%
or just
%%%%%%%%%%%%%%%%%%%%%%%%%%%%%%%%%%%%%%%%%%%%%%%%%%%%%%%%%%%%%%%%%%%
\beq
\mathrm{d}Q^{2} = -j_{\theta} \sin{\!(\theta_{e})}\,\mathrm{d}\theta_{e} ,
\label{eqn_eq:Afan}
\eeq
%%%%%%%%%%%%%%%%%%%%%%%%%%%%%%%%%%%%%%%%%%%%%%%%%%%%%%%%%%%%%%%%%%%
where the transformation Jacobian is represented by
%%%%%%%%%%%%%%%%%%%%%%%%%%%%%%%%%%%%%%%%%%%%%%%%%%%%%%%%%%%%%%%%%%%
\beq
j_{\theta} = - \frac{\sqrt{\lambda_{S}}\,\lambda_{X}^{3/2}}{2M_{d}^{2} \Lb SX - 2m_{e}^{2} \Lb Q^{2} + 2M_{d}^{2} \Rb \Rb} ,
\label{eqn_eq:jtheta}
\eeq
%%%%%%%%%%%%%%%%%%%%%%%%%%%%%%%%%%%%%%%%%%%%%%%%%%%%%%%%%%%%%%%%%%%
and where $\lambda_{X}$ and $X$ are shown below in Eq.~(\ref{eq:eqn_lambdaX0}) and Eq.~(\ref{eq:eqn_rho}), respectively.
Eq.~(\ref{eqn_eq:Afan}) is equivalent to the following relation derived from Eq.~(\ref{eq:eqn_Q2_costheta}):
%%%%%%%%%%%%%%%%%%%%%%%%%%%%%%%%%%%%%%%%%%%%%%%%%%%%
\bea
& & \!\!\!
\mathrm{d}Q^{2} = -\biggl[ \frac{2}{M_{d}} E_{1}E_{2}^{2} - \frac{2}{M_{d}} \frac{\sqrt{E_{1}^{2} - m_{e}^{2}}\,E_{2}^{3}}
{\sqrt{E_{2}^{2} - m_{e}^{2}}}\cos{\!(\theta_{e})} - 
\nonumber \\
& & ~~~~~~~~~
- 2\sqrt{E_{1}^{2} - m_{e}^{2}}\,\sqrt{E_{2}^{2} - m_{e}^{2}} \biggr] \sin{\!(\theta_{e})}\,\mathrm{d}\theta_{e} .
\label{eq:eqn_Jacobian}
\eea
%%%%%%%%%%%%%%%%%%%%%%%%%%%%%%%%%%%%%%%%%%%%%%%%%%%%

The observed cross section as functions of $Q^{2}$ and $\theta_{e}$ for the unpolarized elastic $e-d$ scattering beyond 
ultra-relativistic approximation, including the lowest-order RC contributions, is expressed as follows: 
%%%%%%%%%%%%%%%%%%%%%%%%%%%%%%%%%%%%%%%%%%%%%%%%%%%%%%%%%%%%%%%%%%%
\bea
& & \!\!\!\!
\frac{\mathrm{d}\sigma^{\rm obs}}{\mathrm{d}Q^{2}} = \biggl[ 1 +
\nonumber \\
& & 
+ \frac{\alpha}{\pi} \biggl( \delta_{VR}(Q^{2}) + \delta_{\rm vac}^{l}(Q^{2}) + \delta_{\rm vac}^{h}(Q^{2}) - 
\delta_{\rm inf}(Q^{2}) \biggr) \biggr] \times
\nonumber \\
& & 
\times \left[ e^{(\alpha/\pi)\,\delta_{\rm inf}(Q^{2})} \right] \frac{\mathrm{d}\sigma^{B}}{\mathrm{d}Q^{2}} +
\frac{\mathrm{d}\sigma^{\rm AMM}}{\mathrm{d}Q^{2}} + \frac{\mathrm{d}\sigma_{R}^{F}}{\mathrm{d}Q^{2}} ,
\label{eqn_eq:total_xs}
\eea
%%%%%%%%%%%%%%%%%%%%%%%%%%%%%%%%%%%%%%%%%%%%%%%%%%%%%%%%%%%%%%%%%%%
%%%%%%%%%%%%%%%%%%%%%%%%%%%%%%%%%%%%%%%%%%%%%%%%%%%%%%%%%%%%%%%%%%%
\bea
& & \!\!\!\!
\frac{\mathrm{d}\sigma^{\rm obs}}{\mathrm{d}\theta_{e}} = \biggl[ 1 +
\nonumber \\
& & 
+ \frac{\alpha}{\pi} \biggl( \delta_{VR}(\theta_{e}) + \delta_{\rm vac}^{l}(\theta_{e}) + \delta_{\rm vac}^{h}(\theta_{e}) - 
\delta_{\rm inf}(\theta_{e}) \biggr) \biggr] \times
\nonumber \\
& & 
\times \left[ e^{(\alpha/\pi)\,\delta_{\rm inf}(\theta_{e})} \right] \frac{\mathrm{d}\sigma^{B}}{\mathrm{d}\theta_{e}} +
\frac{\mathrm{d}\sigma^{\rm AMM}}{\mathrm{d}\theta_{e}} + \frac{\mathrm{d}\sigma_{R}^{F}}{\mathrm{d}\theta_{e}} ,
\label{eqn_eq:total_xs_theta}
\eea
%%%%%%%%%%%%%%%%%%%%%%%%%%%%%%%%%%%%%%%%%%%%%%%%%%%%%%%%%%%%%%%%%%%

Let us now describe in details and discuss all the terms in Eq.~(\ref{eqn_eq:total_xs}) and Eq.~(\ref{eqn_eq:total_xs_theta}).

\bigskip
$\bullet~$ {\underline{RC term $\delta_{VR}$}.}
~In order to extract the infrared divergences correctly, one should use the following transformation:
%%%%%%%%%%%%%%%%%%%%%%%%%%%%%%%%%%%%%%%%%%%%%%%%%%%%
\beq
\mathrm{d}\sigma_{R} = \left( \mathrm{d}\sigma_{R} - \mathrm{d}\sigma_{R}^{IR} \right) + \mathrm{d}\sigma_{R}^{IR} = 
\mathrm{d}\sigma_{R}^{F} + \mathrm{d}\sigma_{R}^{IR} ,
\label{eq:eqn_transform}
\eeq
%%%%%%%%%%%%%%%%%%%%%%%%%%%%%%%%%%%%%%%%%%%%%%%%%%%%
where $\sigma_{R}^{F}$ and $\sigma_{R}^{IR}$ on the r.h.s. are the infrared divergence-free and divergence-dependent 
contributions of the cross section, respectively. $\sigma_{R}^{F}$ becomes finite when $k \rightarrow 0$. 
$\sigma_{R}^{IR}$ can be obtained before integration over the variable $\phi_{k}$, as a factorized term in 
front of the Born cross section:
%%%%%%%%%%%%%%%%%%%%%%%%%%%%%%%%%%%%%%%%%%%%%%%%%%%%
\beq
\frac{\mathrm{d}\sigma_{R}^{IR}}{\mathrm{d}Q^{2}} = \frac{1}{R} \lim_{R \rightarrow 0} 
\biggl[ R\,\frac{\mathrm{d}\sigma_{R}}{{\mathrm{d}Q^{2}}} \biggr]
=  -\frac{\alpha}{\pi^{2}}\,\frac{F_{IR}}{R^{2}}\,\frac{\mathrm{d}^{3}k}{k_{0}} 
\frac{\mathrm{d}\sigma^{B}}{\mathrm{d}Q^{2}} ,
\label{eq:eqn_IR}
\eeq
%%%%%%%%%%%%%%%%%%%%%%%%%%%%%%%%%%%%%%%%%%%%%%%%%%%%
where the variable $R$ reads as
%%%%%%%%%%%%%%%%%%%%%%%%%%%%%%%%%%%%%%%%%%%%%%%%%%%%
\beq
R = 2k\!\cdot\!p_{1} = \frac{\upsilon}{1 + \tau} ,
\label{eq:eqn_varR}
\eeq
%%%%%%%%%%%%%%%%%%%%%%%%%%%%%%%%%%%%%%%%%%%%%%%%%%%%
and for the function $F_{IR}$, one should refer to the equations (54), (59) and (60) in \cite{Afanasev:2021nhy}. 

Afterwards, $\mathrm{d}\sigma_{R}^{IR}/\mathrm{d}Q^{2}$ needs to be separated into a soft $\delta_{S}$ and a hard $\delta_{H}$ 
parts by splitting the integration region over the inelasticity $\upsilon$, which can be done by introducing an infinitesimal 
photon energy $\lambda \rightarrow 0$ that is defined in the system $\textit{\bf{p}$_{1}$}$ + $\textit{\bf{q}}$ = 0.
For the formulas of $\delta_{S}$ and $\delta_{H}$, we refer to the equation (62) in \cite{Afanasev:2021nhy}.
For the infrared sum $\delta_{IR} = \delta_{S} + \delta_{H}$, we quote the equation (63) in \cite{Afanasev:2021nhy} and
the equation (26) in \cite{Akushevich:2015toa}.

In order to cancel the infrared divergences, one should also consider the leptonic vertex correction, $\delta_{\rm vert}$, 
in the panel (a) of Fig.~\ref{fig:fig_feynman_rcs}. $\delta_{\rm vert}$ is shown by the equation (41) in \cite{Afanasev:2021nhy},
the equation (36) in \cite{Akushevich:2015toa}, the equation (20) in \cite{Akushevich:1994dn}, and the equation (50) in
\cite{Akushevich:2019mbz}.
Ultimately, the sum of all infrared divergent terms that is now $\delta_{IR} + \delta_{\rm vert}$ 
results in an expression, which is itself free from any infrared divergence. Notably, the components containing the 
infinitesimal photon energy $\lambda$ cancel out explicitly in what is represented as follows:
%%%%%%%%%%%%%%%%%%%%%%%%%%%%%%%%%%%%%%%%%%%%%%%%%%%%%%%%%%%%%%%%%%%
\bea
& & \!\!\!\!
\delta_{VR}(Q^{2}) = \delta_{IR} + \delta_{\rm vert} = 
\nonumber \\
& & ~~~~
= 2 \biggl( \Lb Q^{2} + 2m_{e}^{2} \Rb L_{m} - 1 \biggr) \ln{\!\!\Lb \frac{\upsilon_{\rm cut}}{m_{e} M_{d}} \Rb} +
\nonumber \\
& & ~~~~
+ \frac{1}{2} \Lb SL_{S} + X L_{X} \Rb + S_{\phi}\Lb k_1,k_2,p_2 \Rb +
\nonumber \\
& & ~~~~
+ \Lb \frac{3}{2}Q^{2} + 4m_{e}^{2} \Rb L_{m} - 2 - \frac{(Q^{2} + 2m_{e}^{2})}{\sqrt{\lambda_{m}}} \times
\nonumber \\
& & ~~~~~~~~~~~
\times \Lb \frac{1}{2}\lambda_{m}L_{m}^{2} + 2\,\mbox{Li}_{2}\!\!\Lb \frac{2\sqrt{\lambda_{m}}}{Q^{2} + \sqrt{\lambda_{m}}} \Rb
- \frac{\pi^{2}}{2} \Rb ,
\label{eqn_eq:deltaVR} 
\eea
%%%%%%%%%%%%%%%%%%%%%%%%%%%%%%%%%%%%%%%%%%%%%%%%%%%%%%%%%%%%%%%%%%%
where
%%%%%%%%%%%%%%%%%%%%%%%%%%%%%%%%%%%%%%%%%%%%%%%%%%%%%%%%%%%%%%%%%%%
\bea
 & &\displaystyle
L_{m} = \frac{1}{\sqrt{\lambda_{m}}}\,\ln{\!\!\Lb \frac{\sqrt{\lambda_{m}} + Q^{2}}{\sqrt{\lambda_{m}} - Q^{2}} \Rb} ,
\nonumber \\
& &\displaystyle
L_{S} = \frac{1}{\sqrt{\lambda_{S}}}\,\ln{\!\!\Lb \frac{S + \sqrt{\lambda_{S}}}{S - \sqrt{\lambda_{S}}} \Rb} ,
\nonumber \\
& &\displaystyle
L_{X} = \frac{1}{\sqrt{\lambda_{X}}}\,\ln{\!\!\Lb \frac{X + \sqrt{\lambda_{X}}}
{X-\sqrt{\lambda_{X}}} \Rb} ,
\nonumber \\
& &\displaystyle
\lambda_{X} = X^{2} - 4m_{e}^{2}M_{d}^{2} .
\label{eq:eqn_lambdaX0}
\eea
%%%%%%%%%%%%%%%%%%%%%%%%%%%%%%%%%%%%%%%%%%%%%%%%%%%%%%%%%%%%%%%%%%%
The functional form $S_{\phi}(k_1,k_2,p_2)$ is given by
%%%%%%%%%%%%%%%%%%%%%%%%%%%%%%%%%%%%%%%%%%%%%%%%%%%%%%%%%%%%%%%%%%%
\begin{displaymath}
S_\phi(k_{1},k_{2},p_{2}) = \frac{Q^{2} + 2m_{e}^{2}}{\sqrt{\lambda_m}}\Biggl(\frac 14\lambda_{X}L_{X}^{2} -\frac 14\lambda_{S} L_{S}^{2}\,+
\end{displaymath}
\vskip -0.25truecm
\begin{displaymath}
+ {\rm Li}_{2} \biggl [1 -\frac {(X + \sqrt{\lambda_X})\,T}{8m_{e}^{2}M_{d}^{2}}\biggr ]
+ {\rm Li}_{2} \biggl [1 -\frac {T}{2(X +\sqrt{\lambda_X})}\biggr ] -
\end{displaymath}
\bea
& &\displaystyle ~~~~~~~~~~
- {\rm Li}_{2} \biggl [1 -\frac {Q^{2}(S +\sqrt{\lambda_S})\,T}{2M_{d}^{2}(Q^{2} + \sqrt{\lambda_{m}})^{2}}\biggr ] -
\nonumber\\
& &\displaystyle ~~~~~~~~~~
- {\rm Li}_{2} \biggl [1 -\frac {2m_{e}^{2} Q^{2}T}{(Q^{2} + \sqrt{\lambda_m})^{2}(S + \sqrt{\lambda_{S}})}\biggr ]
\Biggr),
\label{eqn_eq:functionS}
\eea
%%%%%%%%%%%%%%%%%%%%%%%%%%%%%%%%%%%%%%%%%%%%%%%%%%%%%%%%%%%%%%%%%%%
where 
%%%%%%%%%%%%%%%%%%%%%%%%%%%%%%%%%%%%%%%%%%%%%%%%%%%%%%%%%%%%%%%%%%%
\bea
 & &\displaystyle
X = S - Q^{2} ,
\nonumber \\
& &\displaystyle
T = \frac{\Lb Q^{2} + \sqrt{\lambda_m} \Rb \Lb S_{p} - \sqrt{\lambda_m} \Rb}{\sqrt{\lambda_m}} ,
\nonumber \\
& &\displaystyle
S_{p} = S + X = 2S - Q^{2} ,
\label{eq:eqn_rho}
\eea
%%%%%%%%%%%%%%%%%%%%%%%%%%%%%%%%%%%%%%%%%%%%%%%%%%%%%%%%%%%%%%%%%%%
and Li$_{2}$ is Spence's dilogarithmic function
%%%%%%%%%%%%%%%%%%%%%%%%%%%%%%%%%%%%%%%%%%%%%%%%%%%%
\beq
\mbox{Li}_{2}(x) = -\int\limits_{0}^{x} \frac{\ln{\!|1 - y|}}{y}\,\mathrm{d}y .
\label{eqn_eq:functionS10}
\eeq
%%%%%%%%%%%%%%%%%%%%%%%%%%%%%%%%%%%%%%%%%%%%%%%%%%%%

Thus, we have $\delta_{VR}(Q^{2})$ in Eq.~(\ref{eqn_eq:deltaVR}), and the term $\delta_{VR}(\theta_{e})$ is calculated using 
Eq.~(\ref{eq:eqn_Q2_costheta}).

\bigskip
$\bullet~$ {\underline{RC term $\delta_{\rm vac}^{l}$}.} ~In the panel (b) of Fig.~\ref{fig:fig_feynman_rcs}, the diagram 
of vacuum polarization is depicted. In this case, $\delta_{\rm vac}^{l}$ is the leptonic vacuum polarization correction to 
the electric Dirac form factor of the electromagnetic vertex; namely, the polarization originated by $e$, $\mu$, and $\tau$ 
charged leptons:
%%%%%%%%%%%%%%%%%%%%%%%%%%%%%%%%%%%%%%%%%%%%%%%%%%%%%%%%%%%%%%%%%%%
\bea
& & \!\!\!\!
\delta_{\rm vac}^{l}(Q^{2}) = \sum_{i = e,\mu,\tau} \delta_{\rm vac}^{l,i} = 
\sum_{i = e,\mu,\tau} \Lb \frac{2}{3} \Lb Q^{2} + 2m_{i}^{2} \Rb L_{m}^{i} - \right.
\nonumber\\
& & ~~~~~~~~~~~~~~~~~~~~~~~~~~~~~~~~~~~
\left. -\frac{10}{9} + \frac{8m_{i}^{2}}{3Q^{2}} \Lb 1 - 2m_{i}^{2} L_{m}^{i} \Rb \Rb , 
\label{eqn_eq:lept_vac}
\eea
%%%%%%%%%%%%%%%%%%%%%%%%%%%%%%%%%%%%%%%%%%%%%%%%%%%%%%%%%%%%%%%%%%%
where $L_{m}^{i}$ is defined as
%%%%%%%%%%%%%%%%%%%%%%%%%%%%%%%%%%%%%%%%%%%%%%%%%%%%%%%%%%%%%%%%%%%
\beq
L_{m}^{i} = \frac{1}{\sqrt{\lambda_{m}^{i}}}\,\ln{\!\!\Lb \frac{\sqrt{\lambda_{m}^{i}} + Q^{2}}{\sqrt{\lambda_{m}^{i}} - Q^{2}} \Rb} ,
\label{eqn_eq:lept_vac_L}
\eeq
%%%%%%%%%%%%%%%%%%%%%%%%%%%%%%%%%%%%%%%%%%%%%%%%%%%%%%%%%%%%%%%%%%%
with
%%%%%%%%%%%%%%%%%%%%%%%%%%%%%%%%%%%%%%%%%%%%%%%%%%%%%%%%%%%%%%%%%%%
\beq
\lambda_{m}^{i} = Q^{2} \Lb Q^{2} + 4m_{i}^{2} \Rb ,
\label{eqn_eq:lept_vac_lambda}
\eeq
%%%%%%%%%%%%%%%%%%%%%%%%%%%%%%%%%%%%%%%%%%%%%%%%%%%%%%%%%%%%%%%%%%%
being exactly the same as Eq.~(\ref{eqn_eq:lambda_m}) and the first formula of Eq.~(\ref{eq:eqn_lambdaX0}) but both applied
to the three lepton species here.

In this case, we have $\delta_{\rm vac}^{l}(Q^{2})$ in Eq.~(\ref{eqn_eq:lept_vac}), and the term 
$\delta_{\rm vac}^{l}(\theta_{e})$ is calculated using Eq.~(\ref{eq:eqn_Q2_costheta}).

\bigskip
$\bullet~$ {\underline{RC term $\delta_{\rm vac}^{h}$}.} ~$\delta_{\rm vac}^{h}$ is the vacuum polarization correction by hadrons 
to the electric Dirac form factor of the electromagnetic vertex, taken as a fit to experimental cross section data (via dispersion
relations) for $e^{+}e^{-}$ annihilation to hadrons. In our RC framework, we use a parametrization 
\cite{Akushevich:2001yp,Byer:2022bqf}
%%%%%%%%%%%%%%%%%%%%%%%%%%%%%%%%%%%%%%%%%%%%%%%%%%%%%%%%%%%%%%%%%%%
\beq
\delta_{\rm vac}^{h}(Q^{2}) = -\frac{2\pi}{\alpha} \left[ \Delta + \Xi \log\!{(1 + \Sigma\,Q^{2})} \right] ,
\label{eqn_eq:had_vac}
\eeq
%%%%%%%%%%%%%%%%%%%%%%%%%%%%%%%%%%%%%%%%%%%%%%%%%%%%%%%%%%%%%%%%%%%
where the fit parameters $\Delta$, $\Xi$ and $\Sigma$ are shown in Table~\ref{table}. 
%%%%%%%%%%%%%%%%%%%%%%%%%%%%%%%%%%%%%%%%%%%%%
\begin{table}[h!]
\vspace{-3mm}
\centering
  \begin{tabularx}{0.475\textwidth}{|X|X|c|c|}
      \hline
      ~$|t|,~{\rm (GeV/c)^{2}}$ &~~~~~~~~~~~~$\Delta$ &~~$\Xi$ & $\Sigma$ \\
      \hline
    ~~~~~0 $-$ 1    & \!\!$-1.345\times~10^{-9}$     & \!\!$-2.302\times~10^{-3}$   & 4.091\\
      \hline
%    ~~~~~1 - 64   & \!\!$-1.512\times~10^{-3}$     & \!\!$-2.822\times~10^{-3}$   & 1.218 \\
%      \hline
%    ~~~~~$> 64$   & \!\!$-1.344\times~10^{-3}$     & \!\!$-3.068\times~10^{-3}$   & 0.999 \\
%     \hline
  \end{tabularx}
      \caption{The values of the three parameters in Eq.~(\ref{eqn_eq:had_vac}), shown in a given
      range of $|t| = Q^{2}$. The parameters in larger $|t|$ ranges can be seen in Table~1 of \cite{Byer:2022bqf}.}
\vspace{-5mm}
\label{table}
\end{table}
%%%%%%%%%%%%%%%%%%%%%%%%%%%%%%%%%%%%%%%%%%%%%
The hadronic polarization contribution is actually quite small at very low $Q^{2}$ as compared to all-leptonic 
polarization contribution, which is computed in \cite{Arbuzov:2015vba} based on using the Fortran package \texttt{alphaQED} 
of F.~Jegerlehner \cite{Jegerlehner:2011mw}.

We have $\delta_{\rm vac}^{h}(Q^{2})$ in Eq.~(\ref{eqn_eq:had_vac}), and the term $\delta_{\rm vac}^{h}(\theta_{e})$ is
calculated using Eq.~(\ref{eq:eqn_Q2_costheta}).

\bigskip
$\bullet~$ {\underline{RC term $\delta_{\rm inf}$}.} ~As a first approximation, contributions of higher-order RCs are taken 
into consideration by an exponentiation procedure in accordance with \cite{Shumeiko:1978cn} (see also \cite{Akushevich:2015toa,Akushevich:1994dn}). 
This procedure simply accounts for multiple soft photons and corresponding loops for canceling the infrared divergences. 
Then the term designated by $\delta_{\rm inf}$ is used to account for multi-photon radiation at $Q^{2} \rightarrow 0$:
%%%%%%%%%%%%%%%%%%%%%%%%%%%%%%%%%%%%%%%%%%%%%%%%%%%%%%%%%%%%%%%%%%%
\beq
\delta_{\rm inf}(Q^{2}) = \Lb Q_{m}^{2}L_{m} - 1 \Rb \ln{\!\!\Lb \frac{\upsilon_{\rm cut}^{2}}{S(S - Q^{2})} \Rb } .
\label{eqn_eq:delta_inf}
\eeq
%%%%%%%%%%%%%%%%%%%%%%%%%%%%%%%%%%%%%%%%%%%%%%%%%%%%%%%%%%%%%%%%%%%
The exponentiation procedure is manifested by the $e^{(\alpha/\pi) \delta_{\rm inf}}$ term, which stands in the 
observed cross section (see Eq.~(\ref{eqn_eq:total_xs}) and Eq.~(\ref{eqn_eq:total_xs_theta})).

We have $\delta_{\rm inf}(Q^{2})$ in Eq.~(\ref{eqn_eq:delta_inf}), and the term $\delta_{\rm inf}(\theta_{e})$ is 
calculated using Eq.~(\ref{eq:eqn_Q2_costheta}).

\bigskip
$\bullet~$ {\underline{Cross-section terms $\mathrm{d}\sigma^{\rm AMM}/\mathrm{d}Q^{2}$ and $\mathrm{d}\sigma^{\rm AMM}/\mathrm{d}\theta_{e}$}.} 
~There is also the anomalous magnetic moment's contribution to the cross section, which stems from the leptonic vertex correction:
%%%%%%%%%%%%%%%%%%%%%%%%%%%%%%%%%%%%%%%%%%%%%%%%%%%%%%%%%%%%%%%%%%%
\bea
& & \!\!\!
\frac{\mathrm{d}\sigma^{\rm AMM}}{\mathrm{d}Q^{2}} = \frac{\alpha^{3} m_{e}^{2}L_{m}}{2M_{d}^{2}Q^{2}\lambda_{S}} \times
\nonumber \\
& & ~~
\times \biggl( 12M_{d}^{2}\,W_{1d}(Q^{2}) - \Lb Q^{2} + 4M_{d}^{2} \Rb W_{2d}(Q^{2}) \biggr) ,
\label{eqn_eq:sigma_amm}
\eea
%%%%%%%%%%%%%%%%%%%%%%%%%%%%%%%%%%%%%%%%%%%%%%%%%%%%%%%%%%%%%%%%%%%
where the functions $W_{1d}(Q^{2})$ and $W_{2d}(Q^{2})$ are shown in Eq.~(\ref{eqn_eq:app_born4}) and 
Eq.~(\ref{eqn_eq:app_born5}), respectively. The term $\mathrm{d}\sigma^{\rm AMM}/\mathrm{d}\theta_{e}$ 
should be determined from
%%%%%%%%%%%%%%%%%%%%%%%%%%%%%%%%%%%%%%%%%%%%%%%%%%%%%%%%%%%%%%%%%%%
\beq
\frac{\mathrm{d}\sigma^{\rm AMM}}{\mathrm{d}Q^{2}} = -\frac{1}{j_{\theta}\sin{\!(\theta_{e})}}
\frac{\mathrm{d}\sigma^{\rm AMM}}{\mathrm{d}\theta_{e}} ,
\label{eqn_eq:Afan2}
\eeq
%%%%%%%%%%%%%%%%%%%%%%%%%%%%%%%%%%%%%%%%%%%%%%%%%%%%%%%%%%%%%%%%%%%

\bigskip
$\bullet~$ {\underline{Cross-section terms $\mathrm{d}\sigma_{R}^{F}/\mathrm{d}Q^{2}$ and $\mathrm{d}\sigma_{R}^{F}/\mathrm{d}\theta_{e}$}.} 
~The last ingredient of the unpolarized $e-d$ elastic scattering cross section is the infrared-free contribution $d\sigma_{R}^{F}$ 
from Eq.~(\ref{eq:eqn_transform}). In this case, for convenience of calculations of the cross section, the matrix element squared in 
Eq.~(\ref{eqn_eq:brem_xs}) can be represented as
%%%%%%%%%%%%%%%%%%%%%%%%%%%%%%%%%%%%%%%%%%%%%%%%%%%%
\bea
& & \!\!\!\!
\mathcal{M}_{R}^{2} = \frac{(4\pi\alpha)^{3}}{\tilde{Q}^{4}} \times
\nonumber \\
& & ~~~~~~~~~~~~~~
\times L_{R}^{\mu\nu} \Lb \tilde{w}_{\mu\nu}^{1}\,W_{1d}(\tilde{Q}^{2}) + 
\tilde{w}_{\mu\nu}^{2}\,W_{2d}(\tilde{Q}^{2}) \Rb ,
\label{eq:eqn_matrix_sq2}
\eea
%%%%%%%%%%%%%%%%%%%%%%%%%%%%%%%%%%%%%%%%%%%%%%%%%%%%
where the coefficients $w_{\mu\nu}^{i}$ and the functional forms $W_{id}$ are explicitly shown in 
Eqs.~(\ref{eqn_eq:app_born3})-(\ref{eqn_eq:app_born5}). For the radiative leptonic tensor $L_{R}^{\mu\nu}$, see the equations 
(50) and (51) in \cite{Afanasev:2021nhy}. The symbol ``tilde" means that $Q^{2}$ is defined via the shifted $q \rightarrow (q - k)$, 
i.e., by the following replacement of the argument:
%%%%%%%%%%%%%%%%%%%%%%%%%%%%%%%%%%%%%%%%%%%%%%%%%%%%
\beq
\tilde{Q}^{2} = -(q - k)^{2} = Q^{2} + R\tau .
\label{eq:eqn_tilde}
\eeq
%%%%%%%%%%%%%%%%%%%%%%%%%%%%%%%%%%%%%%%%%%%%%%%%%%%%
The tensor contraction $L_{R}^{\mu\nu}\tilde{w}_{\mu\nu}^{i}$ can be expanded in powers of $R$ via a convolution integral that 
is given by the equation (52) in \cite{Afanasev:2021nhy}.

In order to compute $\sigma_{R}^{F}$, one can actually integrate $\sigma_{R}$ over $\phi_{k}$ analytically, and then the integration 
should be performed over the $\tau$ variable\footnote{The integral of $d\sigma_{R}^{F}/d\Omega$ is given by Eq.~(43) in 
\cite{Akushevich:2015toa}.}. After the infrared divergence extraction, the resulting expression is integrated over $\upsilon$. 
Eventually, the finalized finite part of the cross section as a function of $Q^{2}$ reads as
%%%%%%%%%%%%%%%%%%%%%%%%%%%%%%%%%%%%%%%%%%%%%%%%%%%%%%%%%%%%%%%%%%%
\bea
& & \!\!\!\!
\frac{\mathrm{d}\sigma_{R}^{F}}{\mathrm{d}Q^{2}} = -\frac{\alpha^{3}}{2\lambda_{S}} \int\limits_{0}^{\upsilon_{\rm cut}} 
\mathrm{d}\upsilon \sum_{i=1}^{2} \Biggl[ 4\,\frac{J_{0}\,\theta_{B}^{i}\,W_{id}(Q^{2})}{\upsilon\,Q^{4}} +
\nonumber \\
& &
+ \int\limits_{\tau_{q}^{\rm min}}^{\tau_{q}^{\rm max}} \frac{\mathrm{d}\tau}{(1 + \tau)\,\tilde{Q}^{4}}
\sum_{j=1}^{k_{i}} W_{id}(\tilde{Q}^{2})\,R^{j-2}\,\theta_{ij}(\upsilon, \tau, Q^{2}) \Biggr] , 
\label{eqn_eq:sigma_F}
\eea
%%%%%%%%%%%%%%%%%%%%%%%%%%%%%%%%%%%%%%%%%%%%%%%%%%%%%%%%%%%%%%%%%%%
in which $k_{i}$ = \{3,\,4\}. For the three-variable function $\theta_{ij}$, see the equations (53)-(55) in 
\cite{Afanasev:2021nhy}, however, where $M$ must be substituted by $M_{d}$. The limits $\tau_{q}^{\rm min}$ 
and $\tau_{q}^{\rm max}$ at fixed $Q^{2}$ are obtained to be
%%%%%%%%%%%%%%%%%%%%%%%%%%%%%%%%%%%%%%%%%%%%%%%%%%%
\beq
\tau_{q}^{\rm max,min} = \frac{\upsilon + Q^{2} \pm \sqrt{\lambda_{q}}}{2M_{d}^{2}} ,
\label{eqn_eq:taulim}
\eeq
%%%%%%%%%%%%%%%%%%%%%%%%%%%%%%%%%%%%%%%%%%%%%%%%%%%
where $\lambda_{q}$ is given by
%%%%%%%%%%%%%%%%%%%%%%%%%%%%%%%%%%%%%%%%%%%%%%%%%%%
\beq
\lambda_{q} = \Lb \upsilon + Q^{2} \Rb^{2} + 4M_{d}^{2}Q^{2} .
\label{eqn_eq:varY}
\eeq
%%%%%%%%%%%%%%%%%%%%%%%%%%%%%%%%%%%%%%%%%%%%%%%%%%%
Besides, $J_{0}$ is represented by 
%%%%%%%%%%%%%%%%%%%%%%%%%%%%%%%%%%%%%%%%%%%%%%%%%%%%%%%%%%%%%%%%%%%
\beq
J_{0} = 2 \bigl( \Lb Q^{2} + 2m_{e}^{2} \Rb L_{m} - 1 \bigr) ,
\label{eqn_eq:J0}
\eeq
%%%%%%%%%%%%%%%%%%%%%%%%%%%%%%%%%%%%%%%%%%%%%%%%%%%%%%%%%%%%%%%%%%%
and $\theta_{B}^{i}$ is given by Eq.~(\ref{eqn_eq:app_born7}).

In the case of the finite part of the cross section as a function of $\theta_{e}$, we have
%%%%%%%%%%%%%%%%%%%%%%%%%%%%%%%%%%%%%%%%%%%%%%%%%%%%%%%%%%%%%%%%%%%
\bea
& & \!\!\!\!
\frac{\mathrm{d}\sigma_{R}^{F}}{\mathrm{d}\theta_{e}} = \sin{\!(\theta_{e})} \times \Lb \frac{\alpha^{3}}{2\lambda_{S}} \Rb \times
\nonumber \\
& & \!\!\!\!
\times \int\limits_{0}^{\upsilon_{\rm cut}} 
\mathrm{d}\upsilon \sum_{i=1}^{2} \Biggl[ 4j_{\theta}\,\frac{J_{0}\,\theta_{B}^{i}\,W_{id}(Q^{2})}{\upsilon\,Q^{4}} +
\nonumber \\
& & ~~~~~~~~~~~~~~~~~~~
+ J_{\theta}{(\upsilon)} \int\limits_{\tau_{\theta}^{\rm min}}^{\tau_{\theta}^{\rm max}} \frac{\mathrm{d}\tau}{(1 + \tau)\,\tilde{Q}^{4}} \times
\nonumber \\
& & ~~~~~~~~~~~~~~~~~~~
\times \sum_{j=1}^{k_{i}} W_{id}(\tilde{Q}^{2})\,R^{j-2}\,\theta_{ij}\!\!\Lb \upsilon, \tau, Q_{R}^{2}(\upsilon) \Rb \Biggr] ,
\label{eqn_eq:sigma_F_theta}
\eea
%%%%%%%%%%%%%%%%%%%%%%%%%%%%%%%%%%%%%%%%%%%%%%%%%%%%%%%%%%%%%%%%%%%
in which
%%%%%%%%%%%%%%%%%%%%%%%%%%%%%%%%%%%%%%%%%%%%%%%%%%%
\beq
\tau_{\theta}^{\rm max,min} = \frac{\upsilon + Q_{R}^{2}(\upsilon) \pm \sqrt{\lambda_{\upsilon}}}{2M_{d}^{2}} ,
\label{eqn_eq:taulim_theta}
\eeq
%%%%%%%%%%%%%%%%%%%%%%%%%%%%%%%%%%%%%%%%%%%%%%%%%%%
and where 
%%%%%%%%%%%%%%%%%%%%%%%%%%%%%%%%%%%%%%%%%%%%%%%%%%%
\beq
\lambda_{\upsilon} = \Lb \upsilon + Q_{R}^{2}(\upsilon) \Rb^{2} + 4M_{d}^{2} Q_{R}^{2}(\upsilon) ,
\label{eqn_eq:lambda_v}
\eeq
%%%%%%%%%%%%%%%%%%%%%%%%%%%%%%%%%%%%%%%%%%%%%%%%%%%
%%%%%%%%%%%%%%%%%%%%%%%%%%%%%%%%%%%%%%%%%%%%%%%%%%%
\bea
& & \!\!\!\!
Q_{R}^{2}(\upsilon) = \frac{1}{\Lb S + 2M_{d}^{2} \Rb^{2} - \lambda_{S}\cos^{2}{\!(\theta_{e})}} \times
\nonumber \\
& & ~~~
\times \biggl( \Lb S + 2M_{d}^{2} \Rb (\lambda_{S} - \upsilon S) - \lambda_{S} (S - \upsilon) \cos^{2}{\!(\theta_{e})} -
\nonumber \\
& & ~~~~~~~~~~
- 2M_{d} \sqrt{\lambda_{S}}\,\sqrt{\mathcal{D}} \cos{\!(\theta_{e})} \biggr) ,
\label{eqn_eq:QR2}
\eea
%%%%%%%%%%%%%%%%%%%%%%%%%%%%%%%%%%%%%%%%%%%%%%%%%%%
%%%%%%%%%%%%%%%%%%%%%%%%%%%%%%%%%%%%%%%%%%%%%%%%%%%
\bea
& & \!\!\!\!
\mathcal{D} =  M_{d}^{2} \Lb \lambda_{S} + \upsilon (\upsilon - 2S) \Rb - 
\nonumber \\
& & ~~~~~~
- m_{e}^{2}\Lb \lambda_{S} \sin^{2}{\!(\theta_{e})} + 4\upsilon M_{d}^{2} \Rb .
\label{eqn_eq:Dparam}
\eea
%%%%%%%%%%%%%%%%%%%%%%%%%%%%%%%%%%%%%%%%%%%%%%%%%%%
Also,
%%%%%%%%%%%%%%%%%%%%%%%%%%%%%%%%%%%%%%%%%%%%%%%%%%%
\bea
& & \!\!\!\!
J_{\theta}{(\upsilon)} = - \frac{\lambda_{S} - \upsilon S - Q_{R}^{2}(\upsilon)\,(S + 2M_{d}^{2})}
{(S + 2M_{d}^{2})^{2} - \lambda_{S}\cos^{2}{\!(\theta_{e})}} \times
\nonumber \\
& & ~~~~~~~~~~~
\times \biggl( \frac{S + 2M_{d}^{2}}{\cos{\!(\theta_{e})}} + M_{d}\,\sqrt{\frac{\lambda_{S}}{\mathcal{D}}} \Lb S- \upsilon + 2m_{e}^{2} \Rb \biggr) ,
\label{eqn_eq:Jtheta_v}
\eea
%%%%%%%%%%%%%%%%%%%%%%%%%%%%%%%%%%%%%%%%%%%%%%%%%%%
and $j_{\theta} \equiv J_{\theta}{(0)}$.

Again, as in all the previous cases, $Q^{2}$ in Eq.~(\ref{eqn_eq:sigma_F_theta}) should be transformed to $\theta_{e}$ using 
Eq.~(\ref{eq:eqn_Q2_costheta}).

%%%%%%%%%%%%%%%%%%%%%%%%%%%%%%%%%%%%%%%%%%%%%%%%%%%%%%%%%%%
%%%%%%%%%%%%%%%%%%%%%%%%%%%%%%%%%%%%%%%%%%%%%%%%%%%%%%%%%%%
\section{\label{sec:numeric} Numerical results}

In order to produce the figures shown in this section, we have used an event generator for unpolarized elastic $e-d$ scattering 
with hard radiative photons. That generator, a.k.a. DRad e-d is developed recently \cite{Zhou:2023} based on the PRad event generator 
located within the PRad analyzer package \cite{Peng:2017}. This package has already been successfully utilized for monitoring and 
analyzing the PRad experimental data \cite{Xiong:2020kds}, and the event generator there has been built based on the cross sections 
including the $e-p$ and M{\o}ller lowest-order RCs calculated in Ref.~\cite{Akushevich:2015toa}. 

For a consistency check, some of the numerical results produced from the DRad event generator are compared to those produced from 
the MASSRAD package that can be found in \cite{Ilyichev}. In this section, we show several numerical results (for the DRad kinematics)
on the lowest-order RCs, obtained from cross sections and related formulas discussed in the previous sections, as well as in 
Appendix~A and Appendix~B.

%%%%%%%%%%%%%%%%%%%%%%%%%%%%%%%%%%%%%%%%%%%%%%%%%%%%%%%%%%%
\subsection{\label{sec:numeric_rc} The numerical behavior of the unpolarized elastic $e-d$ cross section}

Two inelasticity cut-off values, $\upsilon_{\rm cut}$ and $\upsilon_{\rm min}$, are considered in the calculations here, 
where we have $\upsilon_{\rm min} \leq \upsilon_{\rm cut} \leq \upsilon_{q,\theta}^{\rm max}$, with $\upsilon_{q}^{\rm max}$ 
shown in Eq.~(\ref{eqn_eq:vqmax}), though $\upsilon_{\theta}^{\rm max}$ is represented by the same formula too
\cite{Afanasev:2021nhy}. 
$\upsilon_{\rm cut}$ is an experimental quantity that can be considered as the upper limit 
of Bremsstrahlung integration, being an inelasticity cut-off value for performing calculations in the range of interest. It is 
large enough to account for radiative tail effects. In our calculations, we accept $\upsilon_{\rm cut} = 3.9 ~\rm{GeV^2}$ and 
$\upsilon_{\rm cut}= 8.1 ~\rm{GeV^2}$ that correspond to a $1050 ~\rm{MeV}$ radiative photon at $1.1 ~\rm{GeV}$ beam energy 
and a $2150 ~\rm{MeV}$ radiative photon at $2.2 ~\rm{GeV}$ beam energy, respectively. $\upsilon_{\rm min}$ is an arbitrary 
value for separating the Bremsstrahlung region into soft (non-radiative) and hard (radiative) parts. 
As $\upsilon_{\rm min}$ corresponds to the minimal energy of a radiative photon that can be detected, a requirement 
on $\upsilon_{\rm min}$ must be abided such that it is less than the resolution of the HyCal\footnote{The HyCal is a 
hybrid electromagnetic calorimeter in the PRad experimental setup for detecting energies and scattering angles of 
electrons from both elastic $e-p$ and M{\o}ller ($e-e$) scatterings. An upgraded HyCal has been proposed for deployment 
in the PRad-II \cite{PRad:2020oor} and DRad experimental setups.} (${\rm\Delta}E = 2.6\% \times \sqrt{E_{1}}$, with $E_{1}$ 
given in GeV). The effect of selecting different $\upsilon_{\rm min}$ values is discussed in this section as well.
%Hereby, for obtaining $\mathcal{R}_{ed}$ in Eqs.~(\ref{eqn_eq:ed_ratio_Q2}) and (\ref{eqn_eq:ed_ratio_theta}), 
%the choice of the parameter $\upsilon_{\rm cut}$ should be reasonable, to be within the range of $\upsilon_{q,\theta}^{\rm min}$ and $\upsilon_{q,\theta}^{\rm max}$.
 
\begin{figure}[hbt]
\centering
\includegraphics[width=8.0cm]{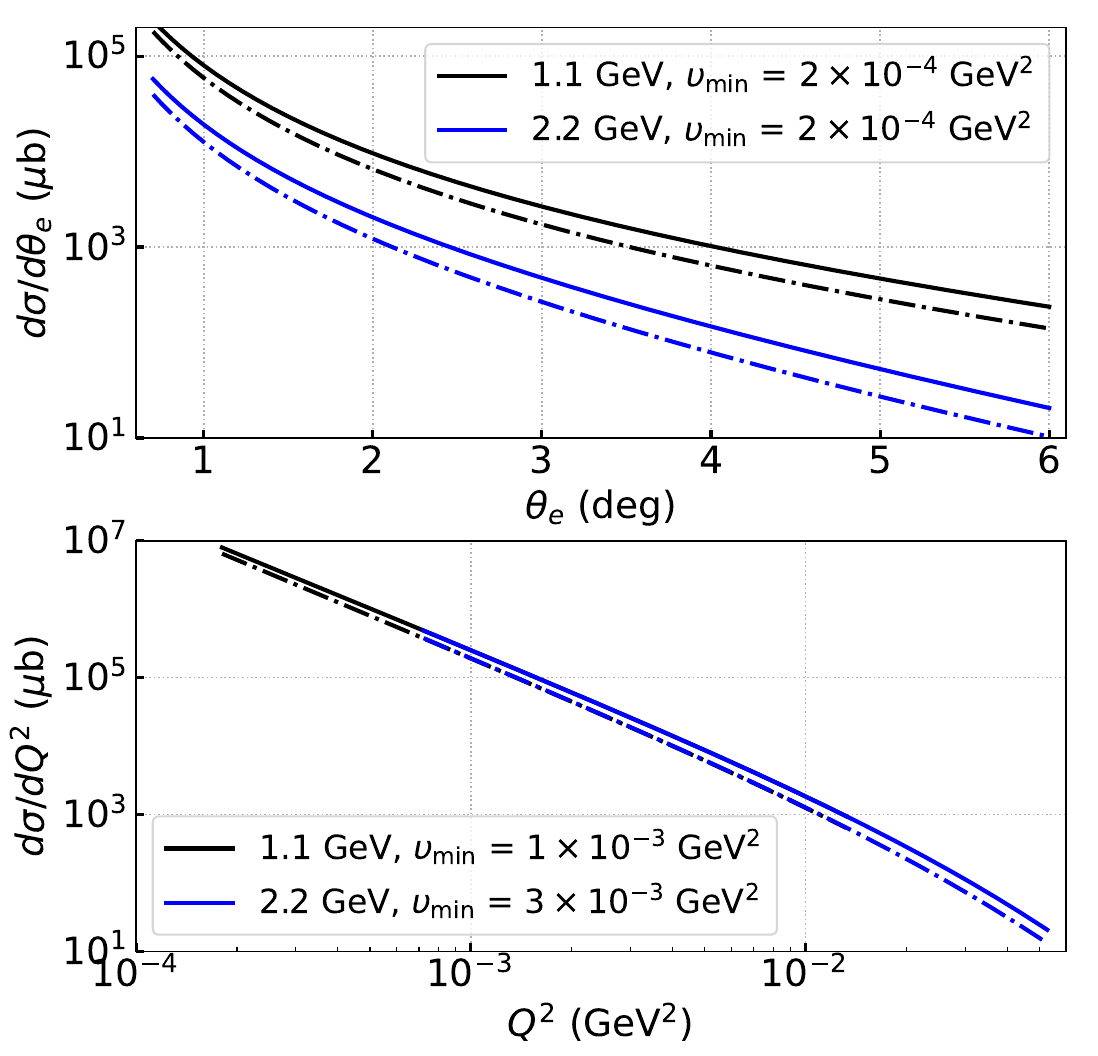}
\vspace{-0.25cm}
\caption{(Color online) 
The observed cross section in the unpolarized elastic $e-d$ scattering as a function of $\theta_{e}$ (top panel) and 
$Q^{2}$ (bottom panel) at $E_{1} = 1.1~{\rm GeV}$ and $2.2~{\rm GeV}$ electron beam energies. The solid lines 
describe the Born cross section calculated from Eqs.~(\ref{eq:eqn_Born}) and (\ref{eqn_eq:dxs_Born_dQ2}). The dot-dashed 
lines describe the cross sections with the soft part of the lowest-order radiative corrections, calculated from 
Eqs.~(\ref{eqn_eq:total_xs_theta}) and (\ref{eqn_eq:total_xs}), and based upon using the Abbott1 form-factor
model in Eq.~(\ref{A1}). 
}
\label{fig:dxs_dtheta_dQ2}
\end{figure}

For the purpose of convenience and effectiveness of numerical simulations, the observed cross section of the 
unpolarized elastic $e-d$ scattering as a function of $Q^{2}$ and $\theta_{e}$ given by Eq.~(\ref{eqn_eq:total_xs}) 
and Eq.~(\ref{eqn_eq:total_xs_theta}) is divided into the aforementioned soft and hard parts in the DRad event 
generator of \cite{Zhou:2023}. The cross section with the soft part of the lowest-order RCs and without any RCs 
is shown in Fig.~\ref{fig:dxs_dtheta_dQ2}. The DRad kinematic range is used for the four-momentum transfer 
squared and the electron scattering angle:
\begin{center}
$Q^{2} = 2\times 10^{-4}~(\gevc)^{2}$~$-$~$5\times 10^{-2}~(\gevc)^{2}$ ,
\end{center}
and
\begin{center}
$\theta_{e} = 0.7^{\circ} - 6.0^{\circ}$ .
\end{center}
%

%https://www.i2pdf.com/crop-pdf
%-----------------------------------------Fig3-----------------------------------------
%---------------------------------------------------------------------------------------

So in Fig.~\ref{fig:dxs_dtheta_dQ2}, the notation $\sigma$ in the vertical axis should be understood as $\sigma^{B}$ and 
$\sigma^{\rm soft}$. The Born cross section (solid lines) is computed from Eq.~(\ref{eq:eqn_Born}) and Eq.~(\ref{eqn_eq:dxs_Born_dQ2}). 
The cross section with the soft RCs is shown as dot-dashed lines. Namely, the cross section 
$\mathrm{d}\sigma^{\rm soft}\big/\mathrm{d}\theta_{e}$ is computed from Eq.~(\ref{eqn_eq:total_xs_theta}) while using 
$\upsilon_{\rm min} = 2\times 10^{-4}~\rm{GeV^2}$. The cross section $\mathrm{d}\sigma^{\rm soft}\big/\mathrm{d}Q^{2}$ 
in turn is computed from Eq.~(\ref{eqn_eq:total_xs}) while the applied $\upsilon_{\rm min}$ is set to be larger because 
of some complications in the numerical integration, used as $\upsilon_{\rm min} = 1\times 10^{-3}~\rm{GeV^2}$ at 
$1.1 ~\rm{GeV}$ and $\upsilon_{\rm min} = 3\times 10^{-3}~\rm{GeV^2}$ at $2.2 ~\rm{GeV}$.

\begin{figure}[hbt!]
\centering
\includegraphics[width=8.0cm]{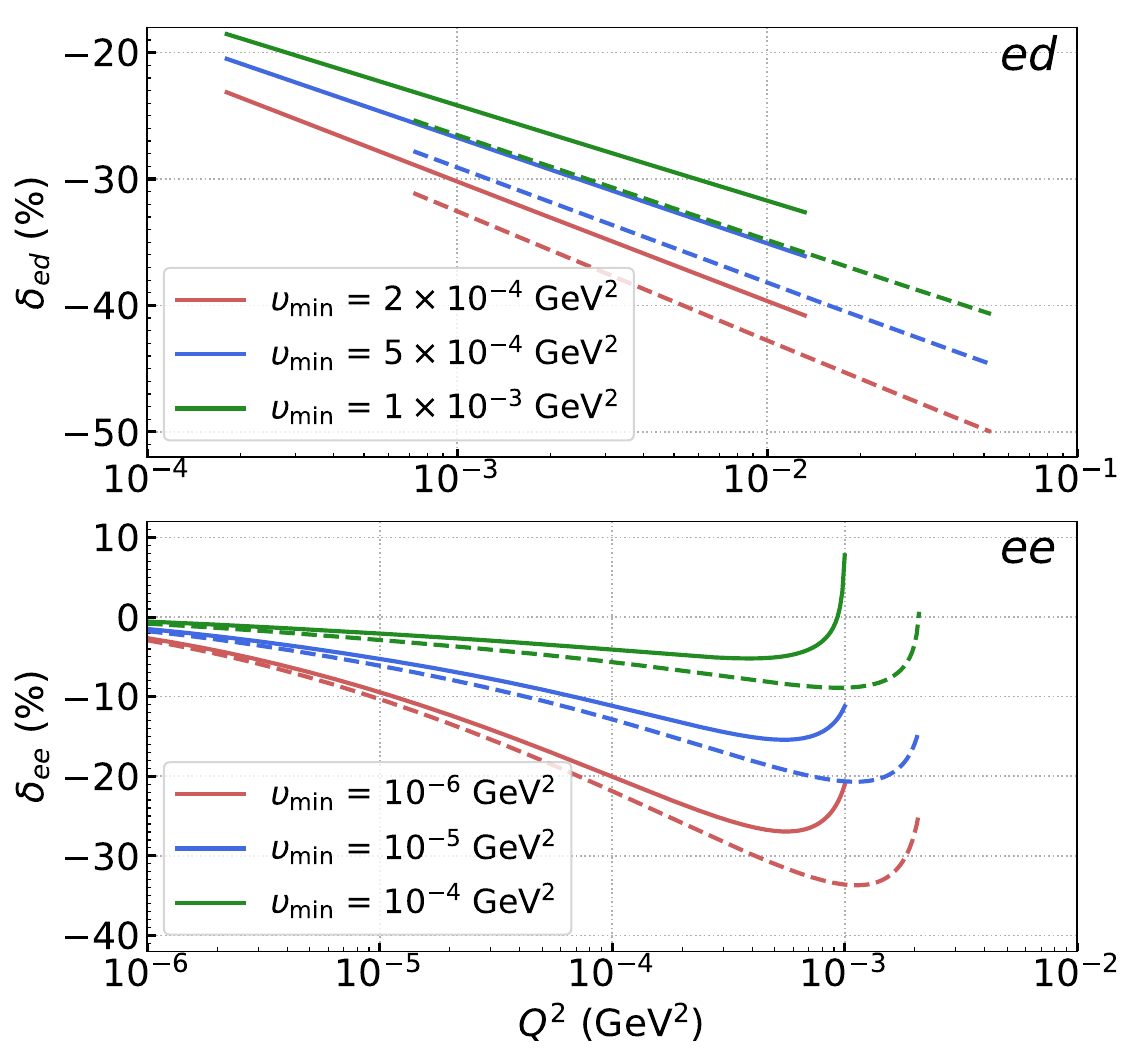}
%\vspace{-0.25cm}
\caption{
(Color online)
(Top panel)
Lowest-order radiative effects in the unpolarized elastic $e-d$ scattering process. The solid curves represent $\delta_{ed}$ 
at $E_{1} = 1.1$ GeV, and the dashed lines represent it at $E_{1} = 2.2$ GeV. The Abbott1 form-factor model in Eq.~(\ref{A1}) 
is used in the calculations of $\delta_{ed}$ curves.
% The DRad kinematic range is used for the four-momentum transfer squared and the electron scattering angle: 
%$Q^{2} = 2\times 10^{-4}~(\gevc)^{2}$ $-$ $5\times 10^{-2}~(\gevc)^{2}$ and $\theta_{e} = 0.7^{\circ} - 6.0^{\circ}$. 
(Bottom panel)
Radiative effects described by $\delta_{ee}$ for the M{\o}ller process, with the same definition. Different ranges for parameters 
and $Q^2$ are chosen according to the kinematics of these two processes within the DRad experimental acceptance. The values of 
$\upsilon_{\rm min}$ and $Q^2$ are chosen according to the kinematics coverage of the DRad experiment for each process.
}
\label{fig:diff_vmin_delta}
\end{figure}

The soft part of the lowest-order RCs as a function of $\theta_{e}$ and $Q^{2}$ can be quantified by the following formulas:
%%%%%%%%%%%%%%%%%%%%%%%%%%%%%%%%%%%%%%%%%%%%%%%%%%%%%%%%%%%%%%%%%%%
\beq
\delta_{ed} = \Lb \frac{\mathrm{d}\sigma^{\rm soft}}{\mathrm{d}\theta_{e}} \bigg/ 
\frac{\mathrm{d}\sigma^{B}}{\mathrm{d}\theta_{e}} \Rb - 1 ,
\label{eqn_eq:ed_ratio_theta}
\eeq
%%%%%%%%%%%%%%%%%%%%%%%%%%%%%%%%%%%%%%%%%%%%%%%%%%%%%%%%%%%%%%%%%%%
and
%%%%%%%%%%%%%%%%%%%%%%%%%%%%%%%%%%%%%%%%%%%%%%%%%%%%%%%%%%%%%%%%%%%
\beq
\delta_{ed} = \Lb \frac{\mathrm{d}\sigma^{\rm soft}}{\mathrm{d}Q^{2}} \bigg/ \frac{\mathrm{d}\sigma^{B}}{\mathrm{d}Q^{2}} \Rb - 1 ,
\label{eqn_eq:ed_ratio_Q2}
\eeq
%%%%%%%%%%%%%%%%%%%%%%%%%%%%%%%%%%%%%%%%%%%%%%%%%%%%%%%%%%%%%%%%%%%
Thereby, $\delta_{ed}$ is defined as the relative difference between the soft and Born differential cross sections.
Then the upper panel in Fig.~\ref{fig:diff_vmin_delta} shows the lowest-order RCs quantified by $\delta_{ed}$ as a function 
of $Q^{2}$ for different values of $\upsilon_{\rm min}$ and for beam energies at $E_{1} = 1.1~{\rm GeV}$ and $2.2~{\rm GeV}$, 
produced using the form-factor Abbott1 model (see Parametrization I~(Abbott1 model) in Appendix~A).
%https://www.i2pdf.com/crop-pdf
%-----------------------------------------Fig4-----------------------------------------
%---------------------------------------------------------------------------------------
%-----------------------------------------Fig5------------------------------------------
%\begin{figure}[hbt!]
%\centering
%\includegraphics[width=8.25cm]{delta_rc_electronmass.pdf}
%\vspace{-0.25cm}
%\caption{
%\color{blue}
%Lowest-order radiative effects in the unpolarized elastic $e-d$ scattering process at $E_{1} = 1.1$ GeV. The solid curves represent $\delta_{ed}$ 
% with $m_{e}$ equal to the standard PDG value ($0.5109989461 \times 10^{-3}$ GeV), and the dashed lines represent $m_{e}$ taking to be smaller 
%values. The Abbott1 form-factor model in Eq.~(\ref{A1}) and $\upsilon_{\rm min} = 1\times 10^{-3}~\rm{GeV^2}$ are used in the calculations of 
%$\delta_{ed}$ curves.
%}
%\label{fig:diff_me_delta}
%\end{figure}
%---------------------------------------------------------------------------------------

Here we emphasize the importance of using the electron mass in the formulas derived 
in Sec.~\ref{sec:rc}. In an extremely low ${Q^{2}}$ range, the ultra-relativistic approximation is not suitable and 
therefore not used in this work. Nonetheless, one can estimate that at ${Q^{2}\sim 10^{-4}~\rm{GeV^2}}$, the term 
$4 m_{e}^{2} \sim 10^{-6}~\rm{GeV^2}$ contributes to 1\% for $\lambda_{m}$ in Eq.~(\ref{eqn_eq:lambda_m}). In turn, 
$\lambda_{m}$ enters in the formula of $L_{m}$ in Eq.~(\ref{eq:eqn_lambdaX0}) (also, in Eq.~(\ref{eqn_eq:deltaVR}) 
or, in Eq.~(\ref{eqn_eq:functionS}) directly) that stands in all the terms in Eq.~(\ref{eqn_eq:total_xs}) 
(and in Eq.~(\ref{eqn_eq:total_xs_theta})) except for $\delta_{\rm vac}^{h}$. 
Corrections to the differential cross section are roughly proportional to $L \equiv \ln\!{(Q^2/m^2_e)}$ (see, e.g, 
Eq.~(\ref{eqn_eq:NLO_corr}) to be discussed a bit later). This $L$ is not the same as $L_{m}$, however, both have 
a similar behavior. One can think of the cross section in Eq.~(\ref{eqn_eq:total_xs}) to behave roughly linearly with  
$L_{m}$. In this case, the effect due to the electron mass in Eq.~(\ref{eqn_eq:total_xs}) 
(or, in Eq.~(\ref{eqn_eq:total_xs_theta})), namely, between the same cross section but beyond and within ultra-relativistic 
approximation, may be considered to be at the level of 1\%.

During the PRad data taking that happened in 2016, the luminosity was monitored by simultaneously measuring the M{\o}ller 
process, and the absolute cross section was normalized to that of the M{\o}ller process to have some level of control of
systematics. The same technique will be utilized in the PRad-II and DRad experiments. In this respect, it is relevant to 
show a similar to $\delta_{ed}$ definition that quantifies the ($Q^{2}$-dependent) M{\o}ller RCs:
%%%%%%%%%%%%%%%%%%%%%%%%%%%%%%%%%%%%%%%%%%%%%%%%%%%%%%%%%%%%%%%%%%%
\beq
\mathcal{\delta}_{ee} = \Lb \frac{\mathrm{d}\sigma_{ee}^{\rm soft}}{\mathrm{d}Q^{2}} \bigg/ 
\frac{\mathrm{d}\sigma_{ee}^{B}}{\mathrm{d}Q^{2}} \Rb - 1 .
\label{eqn_eq:Moller_ratio}
\eeq
%%%%%%%%%%%%%%%%%%%%%%%%%%%%%%%%%%%%%%%%%%%%%%%%%%%%%%%%%%%%%%%%%%%
The bottom plot in Fig.~\ref{fig:diff_vmin_delta} shows examples of this quantity $\delta_{ee}$, which is produced 
using the PRad event generator in \cite{Peng:2017}. 

\begin{figure}[hbt!]
\centering
\includegraphics[width=8.0cm]{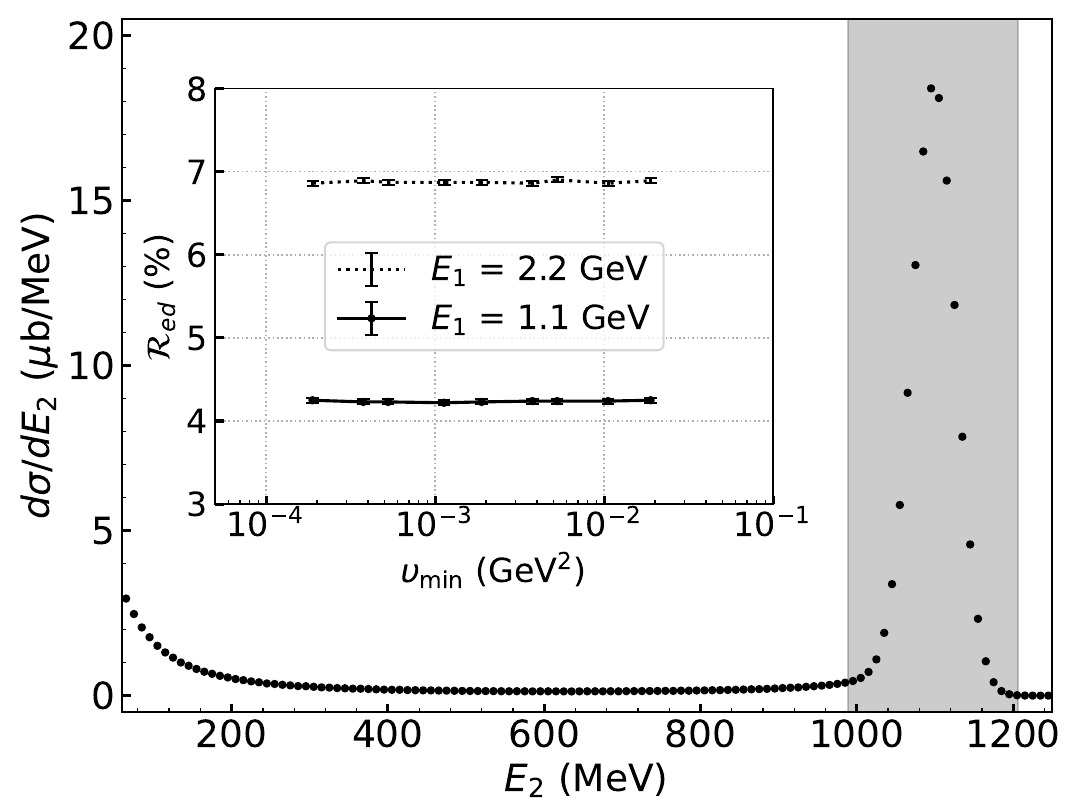}
\vspace{-0.25cm}
\caption{
The cross section $d\sigma/dE_{2}$ for the elastic $e-d$ process at $E_{1} = 1.1$~GeV and within
$0.7^{\circ} < \theta_e < 6.0^{\circ}$. The spectrum includes radiative effects and is smeared by the HyCal 
resolution of ${\rm\Delta}E = 2.6\% \times \sqrt{E_{1}}$. The elasticity cut of 4-$\sigma_E$ is shown by the 
shaded area. The inset shows the experimental radiative effects $\mathcal{R}_{ed}$. With a reasonably small 
$\upsilon_{\rm min}$, the calculated radiative effects under the elastic peak are independent of the choice 
of this parameter.
}
\label{fig:exp_cut}
\end{figure}

As expected, the smaller the value of $\upsilon_{\rm min}$ is selected, the larger the values of $\delta_{ed}$ and 
$\delta_{ee}$ are obtained because $\delta_{ed}$ is dominated by $\delta_{VR}$ (see Eq.~(\ref{eqn_eq:deltaVR})) in 
the soft region. However, only the events within the elastic window will be selected in our future data analysis. 
In this window, the total size of the lowest-order RCs in the unpolarized elastic $e-d$ cross section can be given 
by the following formula:
%%%%%%%%%%%%%%%%%%%%%%%%%%%%%%%%%%%%%%%%%%%%%%%%%%%%%%%%%%%%%%%%%%%
\beq
\mathcal{R}_{ed} = (\sigma^{\rm obs} / \sigma^{B}) - 1 ,
\label{eqn_eq:ed_ratio_total}
\eeq
%%%%%%%%%%%%%%%%%%%%%%%%%%%%%%%%%%%%%%%%%%%%%%%%%%%%%%%%%%%%%%%%%%%
where $\sigma^{\rm obs}$ and $\sigma^{B}$ are integrated cross sections. $\mathcal{R}_{ed}$ is defined as the 
relative difference between the observed and Born cross sections integrated over the DRad acceptance and within 
an energy (elasticity) cut. 
In Fig.~\ref{fig:exp_cut}, we show the normalized distribution of $e-d$ events with hard radiative photons simulated by
the DRad generator at $1.1~{\rm GeV}$. The dashed band describes the elastic $e-d$ region of interest (ROI) from the 
4-$\sigma_E$ energy cut in the HyCal. The inset shows $\mathcal{R}_{ed}$ as a function of $\upsilon_{\rm min}$ 
within the range from $2\times 10^{-4}~\rm{GeV^2}$ to $2\times 10^{-2}~\rm{GeV^2}$. Every 
point is calculated using $10^7$ simulated events. In the ROI, $\mathcal{R}_{ed}$ is very stable (and within the 
statistical uncertainty $0.03\%$), being 4.2\% at $1.1~{\rm GeV}$ and 6.9\% at $2.2~{\rm GeV}$.
%https://www.i2pdf.com/crop-pdf
%-----------------------------------------Fig5-----------------------------------------
%---------------------------------------------------------------------------------------

All the results shown in this Sec.~\ref{sec:numeric_rc} on $e-d$ scattering are generated by using the Abbott1 form-factor 
model. However, the effect coming from the other form-factors models (discussed in Appendix A) is also studied. As shown 
in Fig.~\ref{fig:residual}, the difference in the relative radiative correction due to different form-factor models is 
negligible, being less than $0.01\%$. Although the bin-by-bin observed and Born cross sections due to the other 
three form-factor models differ by up to 0.15\% from the Abbott1 model.

%-----------------------------------------Fig6-----------------------------------------
\begin{figure}[hbt!]
\centering
\includegraphics[width=8.0cm]{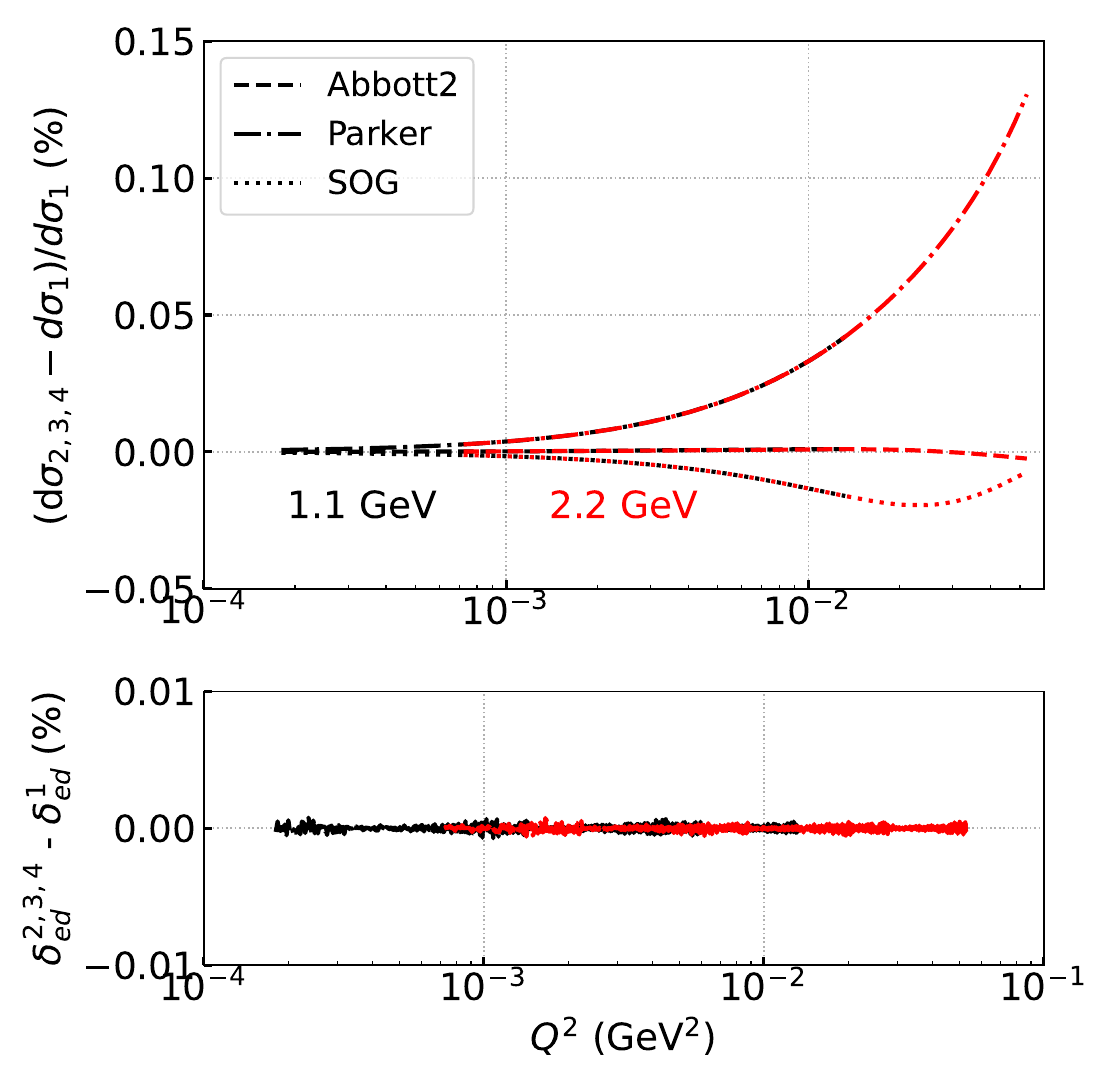}
\vspace{0.0cm}
\caption{
(Top panel) The relative difference of the cross sections between the Abbott2, Parker and SOG form-factor models, 
$d\sigma_{2,3,4}$, and the Abbott1 model, $d\sigma_{1}$. (Bottom panel) The residual of $\delta_{ed}$ between the other 
form-factor models, $\delta_{ed}^{2,3,4}$, and the Abbott1 model, $\delta_{ed}^{1}$. The applied $\upsilon_{\rm min}$ is 
$2\times 10^{-4}~{\rm GeV^{2}}$.
}
\label{fig:residual}
\end{figure} 
%---------------------------------------------------------------------------------------

%\FloatBarrier
%%%%%%%%%%%%%%%%%%%%%%%%%%%%%%%%%%%%%%%%%%%%%%%%%%%%%%%%%%%
\subsection{\label{sec:numeric_rc_syst} Estimation of higher-order RC systematic uncertainty on $r_{d}$ for the DRad experiment}

Ref.~\cite{Arbuzov:2015vba} discusses QED RC corrections to elastic $e-p$ scattering at low energies. In its treatment, 
higher-order corrections relevant to high-precision experiments are presented in an analytic form. The method of
\cite{Arbuzov:2015vba} has been used by PRad but is kind of simplified for estimating the RC systematic uncertainty on the 
measured $r_{p}$. In this section, we use the same procedure but for the elastic $e-d$ scattering to estimate the RC 
systematic uncertainty on $r_{d}$ for the kinematics of the DRad experiment. 

So, as discussed in equation~(2) of \cite{Arbuzov:2015vba}, a typical $\mathcal{O}(\alpha)$ correction to the differential cross section 
is at the magnitude of
%%%%%%%%%%%%%%%%%%%%%%%%%%%%%%%%%%%%%%%%%%%%%%%%%%%%%%%%%%%%%%%%%%%
\bea
& & 
\delta^{(1)} = \frac{\mathrm{d}\sigma^{(1)}}{\mathrm{d}\sigma^{(0)}} \sim \alpha (a \times L + b),
\label{eqn_eq:NLO_corr}
\eea
%%%%%%%%%%%%%%%%%%%%%%%%%%%%%%%%%%%%%%%%%%%%%%%%%%%%%%%%%%%%%%%%%%%
where $\alpha$ is the electromagnetic fine-structure constant, $a$ and $b$ are some constants, $L \equiv \ln\!{(Q^2/m^2_e)}$ 
is the so-called large logarithm. Then the approximate next-to-next-leading order (NNLO) $\mathcal{O}(\alpha^2L^1)$ 
corrections can be calculated by a power counting method. From Eq.~(\ref{eqn_eq:deltaVR}) and 
Eq.~(\ref{eqn_eq:lept_vac}), one can write down
%%%%%%%%%%%%%%%%%%%%%%%%%%%%%%%%%%%%%%%%%%%%%%%%%%%%%%%%%%%%%%%%%%%
\beq
\delta_{VR} + \delta_{\rm vac}^{l} \sim \alpha (a_{VR} \times L + b_{VR}) ,
\label{eqn_eq:approx_deltaVR_vac}
\eeq
%%%%%%%%%%%%%%%%%%%%%%%%%%%%%%%%%%%%%%%%%%%%%%%%%%%%%%%%%%%%%%%%%%%
with 
%%%%%%%%%%%%%%%%%%%%%%%%%%%%%%%%%%%%%%%%%%%%%%%%%%%%%%%%%%%%%%%%%%%
\bea
& & 
a_{VR} = \frac{3}{2} + \frac{2}{3} , 
\nonumber\\
& &
b_{VR} = -2 - \frac{1}{2} \ln{\!\!\Lb \frac{S}{S-Q^2} \Rb}^2 +
\nonumber\\
& & 
~~~~~~~~~~~~
+ \mbox{Li}_{2}\Lb\frac{M_{d}^{2}\,Q^{2}}{S(S - Q^{2})} \Rb - \frac{\pi^{2}}{6} - \frac{10}{9} .
\label{eqn_eq:a_vr}
\eea
%%%%%%%%%%%%%%%%%%%%%%%%%%%%%%%%%%%%%%%%%%%%%%%%%%%%%%%%%%%%%%%%%%%
From Eq.~(\ref{eqn_eq:delta_inf}), we also obtain
%%%%%%%%%%%%%%%%%%%%%%%%%%%%%%%%%%%%%%%%%%%%%%%%%%%%%%%%%%%%%%%%%%%
\beq
\delta_{\rm inf} \sim \alpha (a_{\rm inf} \times L + b_{\rm inf}) ,
\label{eqn_eq:approx_delta_inf}
\eeq
%%%%%%%%%%%%%%%%%%%%%%%%%%%%%%%%%%%%%%%%%%%%%%%%%%%%%%%%%%%%%%%%%%%
with
%%%%%%%%%%%%%%%%%%%%%%%%%%%%%%%%%%%%%%%%%%%%%%%%%%%%%%%%%%%%%%%%%%%
\bea
& &
a_{\rm inf} = \ln{\!\!\Lb \frac{\upsilon_{\rm max}^{2}}{S(S - Q^{2})} \Rb } ,
\nonumber\\
& &
b_{\rm inf} = -\ln{\!\!\Lb \frac{\upsilon_{\rm max}^{2}}{S(S - Q^{2})} \Rb } .
\label{eqn_eq:a_inf}
\eea
%%%%%%%%%%%%%%%%%%%%%%%%%%%%%%%%%%%%%%%%%%%%%%%%%%%%%%%%%%%%%%%%%%%
As a result, the higher-order RCs originate from the interference between the terms in Eq.~(\ref{eqn_eq:approx_deltaVR_vac}) 
and Eq.~(\ref{eqn_eq:approx_delta_inf}):
%%%%%%%%%%%%%%%%%%%%%%%%%%%%%%%%%%%%%%%%%%%%%%%%%%%%%%%%%%%%%%%%%%%
\bea
& & 
\delta^{(2)} = \frac{\mathrm{d}\sigma^{(2)}}{\mathrm{d}\sigma^{(0)}} \sim 
\nonumber\\
& & 
~~~~~~
\sim \alpha^2 L (a_{VR}\times b_{\rm inf} + b_{VR}\times a_{\rm inf} + a_{\rm inf}\times b_{\rm inf}) .
\label{eqn_eq:NNLO}
\eea
%%%%%%%%%%%%%%%%%%%%%%%%%%%%%%%%%%%%%%%%%%%%%%%%%%%%%%%%%%%%%%%%%%%
Note that the term $a_{VR}\times b_{VR}$ is of the order of $\mathcal{O}(\alpha^2L^2)$.
Ultimately, in the DRad kinematic range, we treat the bin-by-bin $\delta^{(2)}$ as the higher-order RC relative 
systematic uncertainty imposed on $r_{d}$, which is
\begin{center}
$0.06\% \sim 0.10\%$ ~~at $E_{1} = 1.1~{\rm GeV}$ beam energy ,
\end{center}
and
\begin{center}
$0.10\% \sim 0.15\%$ ~~at $E_{1} = 2.2~{\rm GeV}$ beam energy .
\end{center}
%

%\FloatBarrier
%%%%%%%%%%%%%%%%%%%%%%%%%%%%%%%%%%%%%%%%%%%%%%%%%%%%%%%%%%%
%%%%%%%%%%%%%%%%%%%%%%%%%%%%%%%%%%%%%%%%%%%%%%%%%%%%%%%%%%%
\section*{\label{sec:sum} Summary and outlook}

In this paper, we presented numerical results from the lowest-order QED radiative correction calculations 
for the unpolarized elastic $e-d$ scattering, by making use of the available four deuteron electromagnetic 
elastic form-factor models. We carried out the calculations within a covariant formalism, whereby the 
derived cross-section formulas may be directly applied to any coordinate system. Besides, we obtained 
our results beyond the ultra-relativistic approximation, which means that the electron mass in not 
neglected in the final expressions. This approximation is no more appropriate in the considered kinematics 
since the calculated RCs are necessary for the deuteron cross-section and charge-radius measurements that
will be accomplished at very low-$Q^{2}$ region in the DRad experiment at Jefferson Lab. As such, we 
also presented our estimation of the bin-by-bin RC relative systematic uncertainty on the deuteron charge 
radius $r_{d}$ due to higher-order effects.

Our framework is anchored upon the Bardin-Shumeiko technique applied to the 
infrared-divergence extraction and cancellation, which has been previously used in
Refs.~\cite{Akushevich:2015toa,Akushevich:1994dn,Akushevich:2001yp,Afanasev:2001jd,Akushevich:2019mbz,Ilyichev:2005rx,Akushevich:2007jc,Afanasev:2021nhy,Afanasev:2022uwm} 
for unpolarized elastic $e-p$ and M{\o}ller cross-section computations with the 
lowest-order RCs, as well as for semi-inclusive DIS cross-section computations again including the 
lowest-order RCs. 

Besides, we do not discuss the hadronic bremsstrahlung and/or vertex correction to the hadronic 
current (deuteron leg) based on the following considerations. 
\begin{itemize}
\item[(i)] First, in our calculations we consider RCs to the leptonic current that include real photon radiation 
from the leptonic leg and vertex, plus also photon self energy. These effects may be calculated without any assumption 
on hadron interactions, and represent the so-called model-independent RC contribution. It is the largest contribution 
to the total RC and can be computed exactly or in the leading-log approximation, if the accuracy provided by this 
approximation is sufficient. By ``exactly'' computing the RCs, we understand analytic expressions obtained without 
any simplified assumption, by having also opportunities for numeric estimates with any predetermined accuracy.

\item[(ii)] Second, the uncertainties of the model-independent RC contribution come only from fits and data 
related to structure functions, whereas the model-dependent corrections (i. e., box-type diagrams, radiation 
by hadronic leg and hadronic vertex) require additional information on the hadron interactions, and therefore 
may contain additional purely theoretical uncertainties, which are not easy to control.

\item[(iii)] Another reason is that $m_{e} \ll M_{d}$, which may result in the hard photon radiation 
probability from the deuteron to be insignificant, thus giving negligible contribution to the total 
cross section.
\end{itemize}

In our approach to the RC treatment, we currently do not consider the two-photon exchange (TPE) box diagrams 
either. The reason is that the hard photon's radiation probability from the deuteron is very low because of $M_{d} \gg m_{e}$.
Also, the form-factor parametrizations we used come from data that had not been corrected for the TPE exchange. 
This means that the TPE effect should be partially included in our results, being nested in the deuteron 
elastic form-factors. Consequently, a possible double counting would be performed in calculations of the box diagram and 
hard photon radiation. By stating ``partially included", we consider the fact that the TPE effects are in 
general non-linear with the scattering-angle parameter $\varepsilon$ (transverse virtual photon polarization), given by
%%%%%%%%%%%%%%%%%%%%%%%%%%%%%%%%%%%%%%%%%%%%%%%%%%%%%%%%%%%%%%%%%%%
\beq
\varepsilon = \frac{1}{1 + 2(1 + \eta)\tan^{2}{(\!\theta/2})} , ~~~~\mbox{with}~~~\eta = \frac{Q^{2}}{4M_{d}^{2}} ,
\label{eqn_eq:epsilon_par}
\eeq
%%%%%%%%%%%%%%%%%%%%%%%%%%%%%%%%%%%%%%%%%%%%%%%%%%%%%%%%%%%%%%%%%%%
while the cross section is linear with respect to $\varepsilon$ for the Born approximation.

Although, important enough pertinent to our case is that the effect of TPE corrections 
have already been studied and determined to be negligible in the kinematics of the PRad experiment. 
Two $e-p$ event generators used in the PRad simulations \cite{Xiong:2020kds} for the purpose of $r_{p}$ 
extraction incorporated the contribution from the TPE processes studied in 
\cite{Tomalak:2018ere,Tomalak:2015aoa,Tomalak:2014sva}. 
That contribution was estimated to be less than $0.2\%$ of the elastic $e-p$ scattering cross section 
in the given PRad kinematic range. Moreover, the cross-section sensitivity to two sets of TPE corrections 
was investigated within the dispersion theoretical framework for elastic $e-p$ and $e^{+}+p$ scattering 
\cite{Lin:2021cnk}, resulting in these corrections to be rather small at the PRad/PRad-II beam energies. 
The TPE corrections for $e-d$ scattering have been discussed in \cite{Gunion:1972bj,Franco:1973uq,Dong:2009gp} 
but in much higher $Q^2$ ranges. As indicated in \cite{Dong:2009gp}, such corrections on the 
deuteron form factors extracted from $e-d$ scattering are expected to be dominated by the TPE effects on 
the nucleon form factors determined from electron-nucleon scattering. Therefore, for DRad one can expect 
a small contribution of TPE as in PRad (at least on the lowest-order RC level), inasmuch as its kinematics 
is very close to that of PRad/PRad-II. 

On the other hand, the TPE corrections in elastic $e-d$ scattering at low $Q^{2}$ could be different from 
those of $e-p$ scattering due to the presence of the deuteron quasielastic breakup channel in the former 
case, which is an intermediate state in the TPE amplitude. Perhaps one needs to compute this contribution, 
including the distortion that arises from the strong interactions in the $pn$ intermediate state close to 
the deuteron threshold \cite{RW:2023}. Ultimately, if it turns out that the TPE corrections 
are not negligible in elastic $e-d$ scattering at low $Q^{2}$ after the TPE theory and/or measurements produce 
new results, then that contribution may be independently added to our present and future cross-section calculations.

It should also be noted that we have only an approximate treatment of higher-order RC effects in our 
current ansatz. Nevertheless, in the near- or mid-term future, it may be important to perform 
state-of-the-art calculations of the NNLO and two-loop irreducible RC contributions for the unpolarized 
elastic $e-d$ scattering process\footnote{Before starting calculations of the NNLO irreducible 
contributions, we may first perform reducible two-loop and quadratic calculations. The quadratic part appears 
in the two-loop correction that stems from the square of one-loop amplitudes \cite{Aleksejevs:2013}.}. 
In that case, the RC component of the total systematic uncertainty on $r_{d}$ could 
be much smaller than what is estimated in Sec.~\ref{sec:numeric_rc_syst}. We currently work on developing 
our approach for such higher-order RC calculations in the unpolarized elastic $e-p$ scattering 
(including the box diagrams too, for computing the TPE corrections), which means that such developments 
may be well applied to the case of $e-d$. Furthermore, similar higher-order RC results and quite 
elaborately-developed frameworks for lepton-proton scattering in Ref.~\cite{Bucoveanu:2018soy} and 
Ref.~\cite{Banerjee:2020rww}, though made for other experiments and kinematics, may likewise be employed
by the PRad collaboration in using them in the DRad experiment \cite{DRad}.

%%%%%%%%%%%%%%%%%%%%%%%%%%%%%%%%%%%%%%%%%%%%%%%%%%%%%%%%%%%
%%%%%%%%%%%%%%%%%%%%%%%%%%%%%%%%%%%%%%%%%%%%%%%%%%%%%%%%%%%
\section*{Acknowledgments}
The work of J.~Z., V.~K., and H.~G. is supported in part by the U.S. Department of Energy, Office of Science, Office of 
Nuclear Physics under contract DE-FG02-03ER41231. The work of C.~P. is supported in part by the same as above but under 
contract DE-AC05-060R23177. S.~S. is supported in part by Brookhaven National Laboratory LDRD 21-045S.
W.~X. is supported by the Shandong Province Natural Science Foundation under grant No. 2023HWYQ-010.

%%%%%%%%%%%%%%%%%%%%%%%%%%%%%%%%%%%%%%%%%%%%%%%%%%%%%%%%%%%
%%%%%%%%%%%%%%%%%%%%%%%%%%%%%%%%%%%%%%%%%%%%%%%%%%%%%%%%%%%
\section*{Data Availability Statement}
This manuscript has no associated data or the data will not be deposited. [Authors' comment: This is a theoretical
study and it has no associated experimental data.]

%%%%%%%%%%%%%%%%%%%%%%%%%%%%%%%%%%%%%%%%%%%%%%%%%%%%%%%%%%%
%%%%%%%%%%%%%%%%%%%%%%%%%%%%%%%%%%%%%%%%%%%%%%%%%%%%%%%%%%%
\appendix
\renewcommand{\theequation}{A\arabic{equation}}
\setcounter{equation}{0}
\section*{Appendix A: Deuteron elastic form-factor models}\label{sec:appA}

Let us now concisely discuss parametrizations describing the available form-factor models: namely Abbott1, 
Abbott2, Parker, and SOG (Sum-of-Gaussian) with given functional forms of $G_{C}^{d}$, $G_{M}^{d}$, and 
$G_{Q}^{d}$\footnote{Ref.~\cite{Zhou:2020cdt} has four of the models discussed but only for $G_{C}^{d}$.}. 
These four models are produced by the fits to the available data in the range from 
$Q^2$ = \(3 \times 10^{-2} ~{\rm (GeV/c)^2}\ {\rm to}\ 1.5~{\rm (GeV/c)^2}\).\\

\noindent {(i) Parametrization I~(Abbott1 model) \cite{Abbott:2000ak}.} \\
In this first parametrization, the three form factors have a generic form given by
%%%%%%%%%%%%%%%%%%%%%%%%%%%%%%%%%%%%%%%%%%%%%%%%%%%%
\bea
& & 
G_{X}^{d}(Q^{2}) = G_{X}^{d}(0) \times
\nonumber \\
& &
~~~~~~~~
\times\left[1 - \left( \frac{Q}{Q^{0}_{X}} \right)^{2}\right]\times\left[1 + \sum_{i=1}^{5} a_{Xi}\,Q^{2i}\right]^{-1} ,
\label{A1}
\eea
%%%%%%%%%%%%%%%%%%%%%%%%%%%%%%%%%%%%%%%%%%%%%%%%%%%%
where $X$ = $C$, $M$, and $Q$. The corresponding three $G_{X}^{d}(0)$ numbers are normalizing factors fixed by the deuteron static 
moments. The free parameters \(Q_{X}^{0}\) and \(a_{Xi}\) can be found on the websites from \cite{Abbott:2000ak,Parker:2020}, 
and are represented as
%%%%%%%%%%%%%%%%%%%%%%%%%%%%%%%%%%%%%%%%%%%%%%%%%%%%
\begin{itemize}
\item[] $Q_{C}^{0} = 4.21~{\rm fm^{-1}}$; \\
$a_{Ci} = 6.740 \cdot 10^{-1}$, ~~$2.246 \cdot 10^{-2}$, ~~$9.806 \cdot 10^{-3}$, ~~$-2.709 \cdot 10^{-4}$, ~~$3.793 \cdot 10^{-6}$; \\

\item[] $Q_{M}^{0} = 7.37~{\rm fm^{-1}}$; \\
$a_{Mi} = 5.804 \cdot 10^{-1}$, ~~$8.701 \cdot 10^{-2}$, ~~$-3.624 \cdot 10^{-3}$, ~~$3.448 \cdot 10^{-4}$, ~~$-2.818 \cdot 10^{-6}$; \\
    
\item[] $Q_{Q}^{0} = 8.10~{\rm fm^{-1}}$; \\
$a_{Qi} = 8.796 \cdot 10^{-1}$, ~~$-5.656 \cdot 10^{-2}$, ~~$1.933 \cdot 10^{-2}$, ~~$-6.734 \cdot 10^{-4}$, ~~$9.438 \cdot 10^{-6}$. \\
\end{itemize}
%%%%%%%%%%%%%%%%%%%%%%%%%%%%%%%%%%%%%%%%%%%%%%%%%%%%

\noindent {(ii) {Parametrization II~(Abbott2 model) \cite{Abbott:2000ak,kobushkin1995deuteron}.} \\
The second parametrization is given by the set of the following expressions:
%%%%%%%%%%%%%%%%%%%%%%%%%%%%%%%%%%%%%%%%%%%%%%%%%%%%
\bea
& & G_{C}^{d}(Q^{2}) = 
\nonumber \\
& &~~~~~~~
= \frac{\Lb G(Q^{2}) \Rb^{2}}{(2\eta + 1)} \left[\left( 1 - \frac{2}{3}\eta \right){g_{00}^+} + \frac{8}{3}\sqrt{2\eta}\,{g_{+0}^+} + 
\right.
\nonumber \\
& &~~~~~~~
~~~~~~~~~~~~~~~~~~~~~~~~ \left. + \frac{2}{3} \left( 2\eta - 1 \right){g_{+ -}^+}\right] ,
\nonumber \\
& & G_{M}^{d}(Q^{2}) = 
\nonumber \\
& &~~~~~~~
= \frac{\Lb G(Q^{2}) \Rb^{2}}{(2\eta + 1)} \left[ 2{g_{00}^+} + \frac{2(2\eta - 1)}{\sqrt{2\eta}}\,{g_{+0}^+} - 
2{g_{+ -}^+} \right] ,
\nonumber \\
& & G_{Q}^{d}(Q^{2}) = 
\nonumber \\
& &~~~~~~~
= \frac{\Lb G(Q^{2}) \Rb^{2}}{(2\eta + 1)} \left[ -{g_{00}^+} + \sqrt{\frac{2}{\eta}}\,{g_{+0}^+} - 
\frac{\eta + 1}{\eta}{g_{+ -}^+} \right] ,
\label{B1}
\eea
%%%%%%%%%%%%%%%%%%%%%%%%%%%%%%%%%%%%%%%%%%%%%%%%%%%%
where $G(Q^2)$ is a dipole form factor of the following form:
%%%%%%%%%%%%%%%%%%%%%%%%%%%%%%%%%%%%%%%%%%%%%%%%%%%%
\beq
G(Q^2) = \left( 1 + \frac{Q^2}{\delta^2} \right)^{-2} ,
\eeq
%%%%%%%%%%%%%%%%%%%%%%%%%%%%%%%%%%%%%%%%%%%%%%%%%%%%
with $\delta$ being a parameter of the order of the nucleon mass $M_{N}$. Besides, we also have
%%%%%%%%%%%%%%%%%%%%%%%%%%%%%%%%%%%%%%%%%%%%%%%%%%%%
\bea
& & {g_{00}^{+}} = \sum_{i=1}^{n}\frac{a_{i}}{\alpha^{2}_{i} + Q^{2}} , \quad {g_{+0}^{+}} = Q \sum_{i=1}^{n}\frac{b_{i}}
{\beta^{2}_{i}+ Q^{2}} , 
\nonumber \\
& &~~~~~~~~~~
\quad {g_{+-}^{+}} = Q^{2} \sum_{i=1}^{n}\frac{c_{i}}{\gamma^2_{i} + Q^{2}} ,
\label{B2}
\eea
%%%%%%%%%%%%%%%%%%%%%%%%%%%%%%%%%%%%%%%%%%%%%%%%%%%%
in which the sets $\left\{ a_{i}, \alpha_{i}^{2} \right\}$, $\left\{ b_{i}, \beta_{i}^{2} \right\}$, and
$\left\{ c_{i}, \gamma_{i}^{2} \right\}$ are fitting parameters. In total, there are twenty-four such parameters, 
which can be found on the website from \cite{Abbott:2000ak}. 

%Those are constrained by the following twelve relations:
%%%%%%%%%%%%%%%%%%%%%%%%%%%%%%%%%%%%%%%%%%%%%%%%%%%%
%\bea
%& &\!\!\!\sum_{i=1}^{n}\frac{a_i}{\alpha_i^{2}} = 1 ,~~\sum_{i=1}^{n} b_{i}  = 0 ,~~~~
%\quad \sum_{i=1}^{n}\frac{b_i}{\beta_i^{2}} = \frac{2 - \mu_{M}^{d}}{2\sqrt{2}M_{d}} ,
%\nonumber \\
%& &\!\!\!\sum_{i=1}^{n} c_{i} = 0 ,~\quad \sum_{i=1}^{n} c_{i}\gamma^{2}_{i} = 0 ,~
%\quad \sum_{i=1}^{n}\frac{c_{i}}{\gamma_{i}^{2}} = \frac{1 - \mu_{M}^{d} - \mu_{Q}^{d}}{4M_{d}^2} ,
%\nonumber \\
%& &\!\!\!\alpha^{2}_{n} = 2M_{d}\,\mu^{(\alpha)} ,~\quad \alpha^{2}_{i} = 
%\alpha^{2}_{1} + \frac{\alpha^{2}_{n} - \alpha^{2}_{1}}{n-1}(i-1) ,
%\nonumber \\
%& &\!\!\!\mbox{for}~\quad i = 1,...,n .
%\label{B3}
%\eea
%%%%%%%%%%%%%%%%%%%%%%%%%%%%%%%%%%%%%%%%%%%%%%%%%%%%
%The last two relations in Eq.~(\ref{B3}) also hold for $\beta_{i}$ and $\gamma_{i}$, where the parameters $\mu^{(\alpha)}$, 
%$\mu^{(\beta)}$, and $\mu^{(\gamma)}$ have the energy dimension, and are of the order of $\Lambda_{QCD} \sim 0.2$~MeV. 
%Eventually, there are twelve free parameters in this model. \\

\noindent {(iii) {Parametrization III~(Parker model) \cite{Parker:2020}.} \\
Based on the re-fits from the first two parametrizations, the third parametrization has in particular some constraints 
for handling the singularities in the functional forms of the $G_{C}^{d}$, $G_{M}^{d}$ and $G_{Q}^{d}$:
%%%%%%%%%%%%%%%%%%%%%%%%%%%%%%%%%%%%%%%%%%%%%%%%%%%%
\bea
& & G_{X}^{d}(Q^{2}) = G_{X}^{d}(0) \times
\nonumber \\
& &
~~~~~~~~ \times\left[1 - \left( \frac{Q}{Q^{0}_{X}} \right)^{2}\right]\times\left[\prod_{i=1}^{5} (1 + |b_{Xi}|\,Q^2) \right]^{-1} ,
\label{Parker}
\eea
%%%%%%%%%%%%%%%%%%%%%%%%%%%%%%%%%%%%%%%%%%%%%%%%%%%%
where $X$ = $C$, $M$, and $Q$, along with $G_{X}^{d}(0)$, $Q_{X}^{0}$ and $b_{Xi}$ that have the same meaning as in Eq.~(\ref{A1}).
The free parameters \(Q_{X}^{0}\) and \(b_{Xi}\) can be found on the websites from \cite{Abbott:2000ak,Parker:2020}, and are 
represented in here as
%%%%%%%%%%%%%%%%%%%%%%%%%%%%%%%%%%%%%%%%%%%%%%%%%%%%
\begin{itemize}
\item[] $Q_{C}^{0} = 4.21~{\rm fm^{-1}}$; \\
$b_{Ci} = 2.709 \cdot 10^{-2}$, ~~$5.451 \cdot 10^{-1}$, ~~$2.709 \cdot 10^{-2}$, ~~$2.708 \cdot 10^{-2}$, ~~$2.709 \cdot 10^{-2}$; \\

\item[] $Q_{M}^{0} = 7.37~{\rm fm^{-1}}$; \\ 
$b_{Mi} = 3.189 \cdot 10^{-2}$, ~~$3.190 \cdot 10^{-2}$, ~~$4.433 \cdot 10^{-1}$, ~~$3.189 \cdot 10^{-2}$, ~~$3.190 \cdot 10^{-2}$; \\

\item[] $Q_{Q}^{0} = 8.10~{\rm fm^{-1}}$; \\
$b_{Qi} = 4.250 \cdot 10^{-2}$, ~~$4.250 \cdot 10^{-2}$, ~~$1.347 \cdot 10^{-3}$, ~~$4.251 \cdot 10^{-2}$, ~~$5.134 \cdot 10^{-1}$. \\
\end{itemize}
%%%%%%%%%%%%%%%%%%%%%%%%%%%%%%%%%%%%%%%%%%%%%%%%%%%%

\noindent {(iv) {Parametrization IV~(SOG model) \cite{Abbott:2000ak,Sick:1974suq,Zhou:2020}.} \\
The fourth parametrization is obtained using the SOG procedure, by which the final generic form of the three form 
factors read as
%%%%%%%%%%%%%%%%%%%%%%%%%%%%%%%%%%%%%%%%%%%%%%%%%%%%
\bea
& & G_{X}^{d}(Q^{2}) = 
\nonumber \\
& &
= G_{X}^{d}(0)\times e^{-\frac{1}{4}Q^2 \gamma^2}\times \sum_{i=1}^{N} \frac{A_{Xi}}{1 + \Lb 2R_{Xi}^{2}/\gamma^{2} \Rb} \times
\nonumber \\
& &~~~~~~~
\times \left[\cos{\!(Q\,R_{Xi})} + \frac{2R_{Xi}^{2}}{\gamma^2}\frac{\sin{\!(Q\,R_{Xi})}}{Q\,R_{Xi}}\right] .
\label{SOG}
\eea
%%%%%%%%%%%%%%%%%%%%%%%%%%%%%%%%%%%%%%%%%%%%%%%%%%%%
where we again have $X$ = $C$, $M$, and $Q$. In the coordinate space, in which the parametrization in Eq.~(\ref{SOG}) 
is described better, it corresponds to the $\rho(R)$ density profile\footnote{The density $\rho$ is a function of the 
distance $R$ between the nucleons and the deuteron center of mass.}, which is given in terms of a Gaussian sum located 
at arbitrary radius $R_{Xi}$, with amplitudes $A_{Xi}$ fitted to the three form-factor data sets, given also the fixed
Gaussian width $\gamma = 0.8\!\cdot\!\sqrt{2/3}$~fm.
In our fitting procedure, we accept $N = 12$. There are free fitting eleven parameters: namely, ten Gaussian amplitudes 
$\left\{ \right.$$A_{X1}$, $A_{X2}$, ..., $A_{X10}$$\left. \right\}$ at ten points of $\left\{ \right.$$R_{X1}$, 
$R_{X2}$, ..., $R_{X10}$$\left. \right\}$ $< 4~{\rm fm}$, respectively. Besides, there is one overall amplitude $A_{X11}$ 
corresponding to another $R_{X11}$ point that is located in the range from $4~{\rm fm}$ to $10~{\rm fm}$. For determining 
the normalization, there is one more amplitude $A_{X12,\rm norm}$ taken at the point of $R_{X12,\rm norm} = 0.4~{\rm fm}$. 
All the given amplitudes satisfy the condition of $\sum\limits_{i=1}^{12} A_{Xi} = 1$. 

Thereby, to find the parameters $A_{Xi}$, we first randomly generate a set of $R_{Xi}$ in the entire range mentioned above, 
then fit the functional forms in Eq.~(\ref{SOG}) to the  $G_{X}^{d}$ data sets from Table~1 of \cite{Abbott:2000ak} 
(see also \cite{Zhou:2020}). The sets of $R_{Xi}$ values are generated repeatedly until the determined $\chi^2$ value gets 
minimized and converged. With the eleven fixed $R_{Xi}$ and eleven free parameters $A_{Xi}$, the final fits are obtained 
\cite{Zhou:2020} to be $\chi^2/{\rm NDF} \simeq 1.625$ for $G_{C}^{d}$, $\chi^2/{\rm NDF} \simeq 0.094$ for $G_{M}^{d}$,
$\chi^2/{\rm NDF} \simeq 1.371$ for $G_{Q}^{d}$.
In this regard, let us represent below all the form-factor fitting parameters and $R_{Xi}$ values in the SOG model.
%%%%%%%%%%%%%%%%%%%%%%%%%%%%%%%%%%%%%%%%%%%%%%%%%%%%
\begin{itemize}
\item[] $Q_{C}^{0} = 4.21~{\rm fm}$; \\
$A_{C1,...,C11} = 0.072$, ~~236.436, ~~1454.11, ~~59.132, ~~-1052.67, ~~-17.789, ~~-8.492, ~~995.005, ~~-1323.4, ~~-365.277, ~~23.591; \\
$A_{C12}=1-\sum_{i=1}^{11} A_{Ci}$; \\
$r_{1,...,10} = 2.706$, ~~$3.911$, ~~$3.831$, ~~$1.490$, ~~$2.468$, ~~$2.838$, ~~$1.493$, ~~$1.562$, ~~$3.863$, ~~$3.755~{\rm fm}$; \\ 
$r_{11} = 5.197~{\rm fm}$; \\
$r_{12} = 4.000 \cdot 10^{-1}~{\rm fm}$; \\

\item[] $Q_{M}^{0} = 7.37~{\rm fm}$; \\
$A_{M1,...,M11} = -0.079$, ~~-17.805, ~~1.019, ~~-22.073, ~~3.337, ~~-147.885, ~~2.727, ~~-5.093, ~~28.676, ~~0.281, ~~158.688;\\
$A_{M12}=1-\sum_{i=1}^{11} A_{Mi}$; \\
$r_{1,...,10} = 3.314$, ~~$4.413 \cdot 10^{-2}$, ~~$3.688$, ~~$2.089$, ~~$2.332$, ~~$3.341$, ~~$1.955$, ~~$3.725$, ~~$5.339 \cdot 10^{-1}$, ~~$3.160~{\rm fm}$; \\
$r_{11} = 7.008~{\rm fm}$; \\
$r_{12} = 4.000 \cdot 10^{-1}~{\rm fm}$; \\

\item[] $Q_{Q}^{0} = 8.10~{\rm fm}$; \\
$A_{Q1,...,Q11} = 0.359$, ~~3.221, ~~0.445, ~~-9.954, ~~-7.354, ~~-247.766, ~~14.768, ~~-302.738, ~~548.153, ~~1.609, ~~-0.261;\\
$A_{Q12}=1-\sum_{i=1}^{11} A_{Qi}$; \\
$r_{1,...,10} = 5.702 \cdot 10^{-1}$, ~~$3.987$, ~~$2.250$, ~~$2.274$, ~~$3.596$, ~~$2.222$, ~~$3.420$, ~~$3.731$, ~~$1.363$, 
~~$2.778~{\rm fm}$; \\
$r_{11} = 9.511~{\rm fm}$; \\
$r_{12} = 4.000 \cdot 10^{-1}~{\rm fm}$.
\end{itemize}
%%%%%%%%%%%%%%%%%%%%%%%%%%%%%%%%%%%%%%%%%%%%%%%%%%%%

%%%%%%%%%%%%%%%%%%%%%%%%%%%%%%%%%%%%%%%%%%%%%%%%%%%%%%%%%%%
%%%%%%%%%%%%%%%%%%%%%%%%%%%%%%%%%%%%%%%%%%%%%%%%%%%%%%%%%%%
\appendix
\renewcommand{\theequation}{B\arabic{equation}}
\setcounter{equation}{0}
\section*{Appendix B: Born cross section in the ansatz of \cite{Afanasev:2021nhy}}\label{sec:appB}

In this appendix, following Refs.~\cite{Afanasev:2021nhy} and \cite{Gakh:2018sat}, we represent the Born cross section 
of the unpolarized elastic $e-d$ scattering cross section (see Fig.~\ref{fig:fig_feynman_born}) in the nomenclature of 
Sec.~\ref{sec:xs}.

The matrix element squared expressed through the convolution of the hadronic and leptonic tensors is given by
%%%%%%%%%%%%%%%%%%%%%%%%%%%%%%%%%%%%%%%%%%%%%%%%%%%%%%%%%%%%%%%%%%%
\beq
\mathcal{M}_{B}^{2} = 16\pi^{2} \frac{\alpha^{2}}{Q^{4}}\,W_{\mu\nu}(q)\,L^{\mu\nu} .
\label{eqn_eq:app_born1}
\eeq
%%%%%%%%%%%%%%%%%%%%%%%%%%%%%%%%%%%%%%%%%%%%%%%%%%%%%%%%%%%%%%%%%%%
The leptonic tensor averaged over the incident unpolarized electron spin and summed over polarizations of the 
scattered electron has the form of 
%%%%%%%%%%%%%%%%%%%%%%%%%%%%%%%%%%%%%%%%%%%%%%%%%%%%%%%%%%%%%%%%%%%
\beq
L^{\mu\nu} = q^{2}g^{\mu\nu} + 2 \Lb k^{1\mu}k^{2\nu} + k^{1\nu}k^{2\mu} \Rb .
\label{eqn_eq:app_born2}
\eeq
%%%%%%%%%%%%%%%%%%%%%%%%%%%%%%%%%%%%%%%%%%%%%%%%%%%%%%%%%%%%%%%%%%%
The hadronic tensor for the unpolarized target and recoil deuterons can be rearranged into the standard covariant 
form as follows:
%%%%%%%%%%%%%%%%%%%%%%%%%%%%%%%%%%%%%%%%%%%%%%%%%%%%%%%%%%%%%%%%%%%
\bea
& & W_{\mu\nu}(q) = - \biggl( g_{\mu\nu} - \frac{q_{\mu}q_{\nu}}{q^{2}} \biggr) W_{1d}(-q^{2}) +
\nonumber \\
& &~~~~~~~
+ \frac{1}{M_{d}^{2}} \biggl( p_{1\mu} + \frac{q_{\mu}}{2} \biggr) \biggl( p_{1\nu} + \frac{q_{\nu}}{2} \biggr) W_{2d}(-q^{2}) =
\nonumber \\
& &~~~~~~~~~~~
= \sum_{i = 1}^{2} w_{\mu\nu}^{i}(q)\,W_{id}(-q^{2}) ,
\label{eqn_eq:app_born3}
\eea
%%%%%%%%%%%%%%%%%%%%%%%%%%%%%%%%%%%%%%%%%%%%%%%%%%%%%%%%%%%%%%%%%%%
where $W_{1d}(-q^{2})$ and $W_{2d}(-q^{2})$ are directly related to the unpolarized elastic structure functions $B_{d}(Q^{2})$ 
and $A_{d}(Q^{2})$, respectively, as shown in Eq.~(\ref{eq:eqn_deuteronstrucfunc}):
%%%%%%%%%%%%%%%%%%%%%%%%%%%%%%%%%%%%%%%%%%%%%%%%%%%%%%%%%%%%%%%%%%%
\beq
W_{1d}(Q^{2}) = 2M_{d}^{2}\,B_{d}(Q^{2}) ,
\label{eqn_eq:app_born4}
\eeq
%%%%%%%%%%%%%%%%%%%%%%%%%%%%%%%%%%%%%%%%%%%%%%%%%%%%%%%%%%%%%%%%%%%
and
%%%%%%%%%%%%%%%%%%%%%%%%%%%%%%%%%%%%%%%%%%%%%%%%%%%%%%%%%%%%%%%%%%%
\beq
W_{2d}(Q^{2}) = 4M_{d}^{2}\,A_{d}(Q^{2}) .
\label{eqn_eq:app_born5}
\eeq
%%%%%%%%%%%%%%%%%%%%%%%%%%%%%%%%%%%%%%%%%%%%%%%%%%%%%%%%%%%%%%%%%%%
Ultimately, after computing the tensor convolution, we can write down the formula of the Born cross section as a function 
of $Q^{2}$ in the one-photon exchange approximation, in the given reference system with the target deuteron at rest:
%%%%%%%%%%%%%%%%%%%%%%%%%%%%%%%%%%%%%%%%%%%%%%%%%%%%%%%%%%%%%%%%%%%
\bea
& & \!\!\!\!
\frac{\mathrm{d}\sigma^{B}}{\mathrm{d}Q^{2}}\!\Lb E_{1}, Q^{2} \Rb  = 
\nonumber \\
& &~~~~~~~~~~~~~~
= \frac{2\pi\alpha^{2}}{\lambda_{S}\,Q^{4}} \biggl( \theta_{B}^{1}\,W_{1d}(Q^{2}) + \theta_{B}^{2}\,W_{2d}(Q^{2}) \biggl) ,
\label{eqn_eq:app_born6}
\eea
%%%%%%%%%%%%%%%%%%%%%%%%%%%%%%%%%%%%%%%%%%%%%%%%%%%%%%%%%%%%%%%%%%%
with
%%%%%%%%%%%%%%%%%%%%%%%%%%%%%%%%%%%%%%%%%%%%%%%%%%%%%%%%%%%%%%%%%%%
\beq
\theta_{B}^{1} = Q^{2} - 2m_{e}^{2} ,~~~~~\theta_{B}^{2} = \frac{S X - M_{d}^{2}Q^{2}}{2M_{d}^{2}} .
\label{eqn_eq:app_born7}
\eeq
%%%%%%%%%%%%%%%%%%%%%%%%%%%%%%%%%%%%%%%%%%%%%%%%%%%%%%%%%%%%%%%%%%%

%%%%%%%%%%%%%%%%%%%%%%%%%%%%%%%%%%%%%%%%%%%%%%%%%%%%%%%%%%%
%%%%%%%%%%%%%%%%%%%%%%%%%%%%%%%%%%%%%%%%%%%%%%%%%%%%%%%%%%%

\end{document}